\documentclass[11pt,a4paper]{article}

 \usepackage{hhline}
 \usepackage{supertabular}
\usepackage{amssymb}
\usepackage{graphicx}    
\usepackage{amsmath}
\usepackage{tikz}
\usepackage{tabularx}
\usepackage{longtable}
\usepackage{ltxtable}
\usepackage{float}

\usepgflibrary{arrows}
\usetikzlibrary{shadows}
\usepackage{url}
\newcommand{\ndt}[1]{$#1$}
\newcommand{\ndp}[1]{$#1$}

\begin{document}

\clearpage %\pagestyle{Standard}
\begin{tabular}{l m{5cm} r}
\includegraphics[width=2.858cm,height=2.313cm]{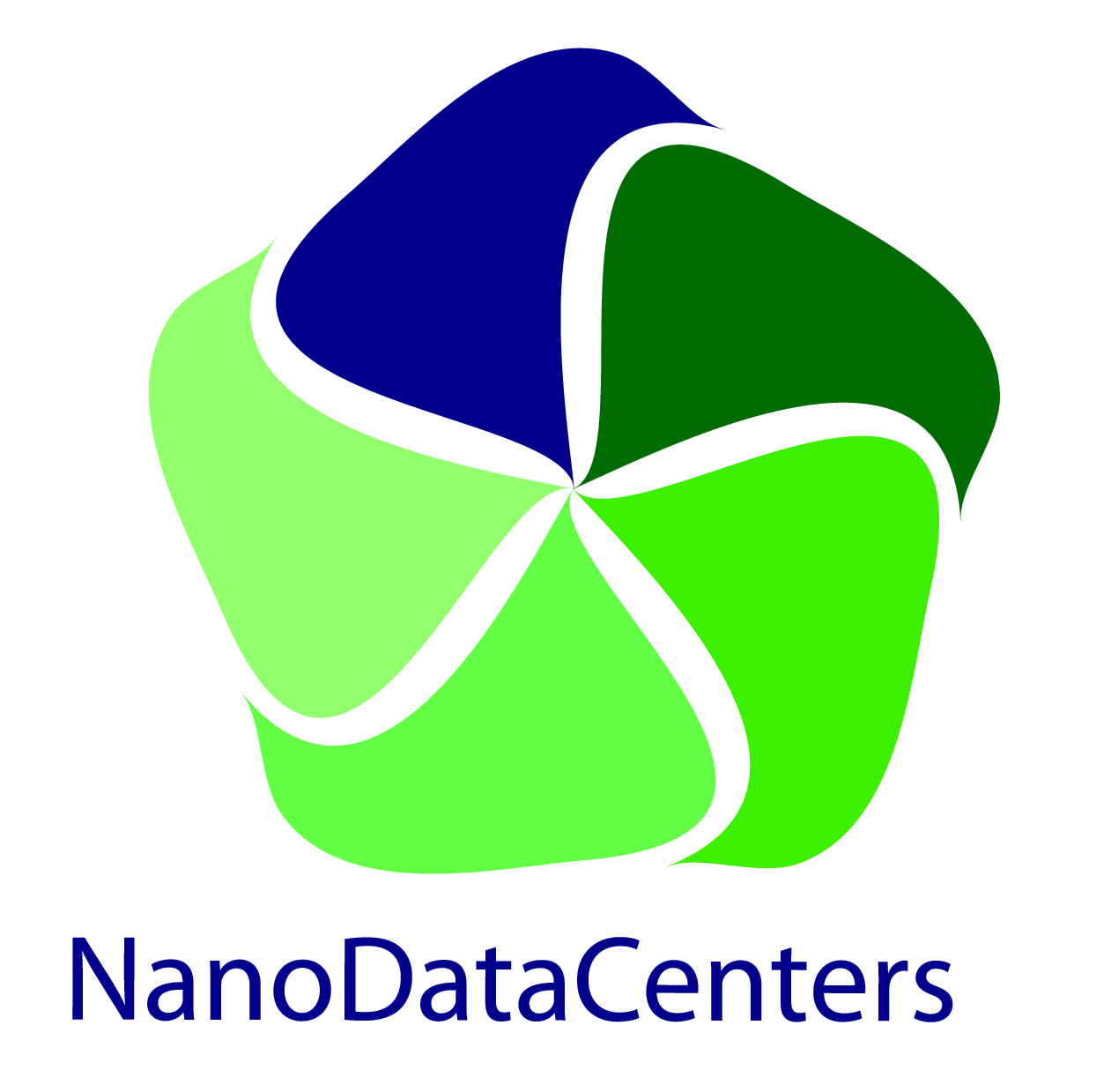}&\hspace{5cm}&
\includegraphics[width=2.864cm,height=2.827cm]{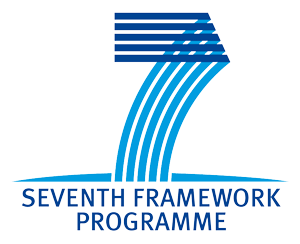}\\
\end{tabular}
\bigskip
\bigskip
% \bigskip
% \bigskip
% \bigskip

{\centering\bfseries
Project no. 223850
\par}
\bigskip
{\centering\bfseries
NANODATACENTERS
\par}
\bigskip
% \bigskip
% \bigskip
{\centering
\textbf{Deliverable }\textbf{D3.2: Final Architecture
Specification of security,
privacy, and incentive
mechanisms}%\textbf{Design and Decomposition}
\par}
\bigskip
{\centering
Due date of deliverable: 31\textsuperscript{st} October 2009
\par}
{\centering
Submission date: 6\textsuperscript{th} November 2009
\par}

\bigskip

\bigskip
\ \\
\begin{tabular}{l l}
Instrument & STREP \\
Start date of project & May 1\textsuperscript{st} 2008 \\
Duration & 36 months \\
\begin{minipage}{4cm}
Organisation name of lead contractor for this deliverable: 
\end{minipage}& SIT \\
Revision & v1.0 \\
Authors & \begin{minipage}{8cm}
          Nicolai Kuntze, J\"urgen Repp, Hervais Simo Fhom, Andreas Fuchs, Ine-Saf Benaissa
          \end{minipage}\\
\end{tabular}

\bigskip

\bigskip
\ \\
{\bfseries
Abstract}
\ \\
In this document, we define the NADA security architecture based on refined use case scenarios, 
a derived high level model and security analysis. For the architecure design and verification we are applying
the well known STRIDE model.

\begin{flushleft}
%\tablehead{}
\centering
\begin{tabular}{|m{1cm}|m{12cm}|m{1cm}|}
\hline
\multicolumn{3}{|c|}{\begin{minipage}{14cm}\bfseries Project co-funded by the European Commission within the Seventh Framework Programme \end{minipage}}\\
\hline
\multicolumn{3}{|c|}{\bfseries Dissemination Level}\\
\hline
\bfseries PU & Public & \checkmark \\
\hline
\bfseries PP & Restricted to other programme participants (including the Commission Services) & \\
\hline
\bfseries RE & Restricted to a group specified by the consortium (including the Commission Services) &\\
\hline
\bfseries CO & Confidential, only for members of the consortium (including the Commission Services) &\\
\hline
\end{tabular}
\end{flushleft}

\bigskip

\newpage

\tableofcontents
\newpage

\listoffigures

\listoftables

\newpage

%\mainmatter  % start of an individual contribution

% first the title is needed
% \title{D3.2 - Final Architecture
% Specification of security,
% privacy, and incentive
% mechanisms}
% 
% 
% 
% \maketitle 

%\begin{abstract}

%\end{abstract}
\section{Introduction}\label{sec:Introduction}

%TODO Introduction to the Document, 
Distribution of virtual goods over the IP based infrastructure offered by the
Internet requires efficient techniques w.r.t.\ to the utilization of the existing
resources. One approach here applies methods from the peer to peer domain to
ensure that the required traffic mostly is situated in the cost efficient last
mile. This paper presents this approach and discusses the security implications.

The proposed security and trust architecture includes solutions for integrity
protection of data as well as for software on the device, exclusion of
manipulated nodes from the network, and isolation between owned applications by
different stakeholders residing in parallel on the same platform. All solutions
can be build on existing secure hardware anchors as provided by the Trusted
Platform Module (TPM) and its certification infrastructure.
This document presents 
\begin{description}
\item[Use Case Scenarios] The scenarios are defining the operations of the
system that are to be covered by the resulting architecture. The scenarios are inputs for the threat analysis and provide the reference of the security architecture.
\item[High level Model] Derived from the scenarios given in this document and
D2.1 high level scenarios are defined introducing the stakeholders, entities, and
interactions between them. This document provides an update to D3.1 \cite{D31} as it is 
included in the joint deliverable D1.1-D3.1 \cite{D11}\cite{D31}.  
\item[Threat Analysis] Given the scenarios the threat analysis provides threats to be covered by the security analysis and therefore also a reference for the evaluation planed to be done in D3.4. \cite{D34}.
\item[Security Architecture] The resulting security architecture is presented in a
threat model according to the STRIDE model. 
\end{description}

\section{Use Scenarios}\label{sec:Use Scenarios}
For the NaDa security architecture the following use scenarios are considered. The selection
of the use scenarios was done according to D1.1 \cite{D11}\cite{D31} supporting the basic operations
required to run the NaDa system. Deployment of nodes and slices are the basic
operations required to establish the service at the node side. The handling
of user requests provides basic functionality towards the end user. Monitoring
provides essential information required for control and QoE enforcement.

This chapter first introduces the used terms and NaDa primitives to provide for a common
definition of vocabulary in the security model and architecture. After this the
use scenarios are introduced. 
%TODO Introduction what use scenarios are given

\subsection{Used Terms}\label{sec:Used Terms}

This subsection explains terms used to describe security relevant control messages in the NaDa security architecture.

\begin{description}
 \item[\ndt{APP\_ID}] Identifier for a certain customer application which is running
  in slice assigned by the customer itself.
 \item[\ndt{Customer\_ID}] Unambiguously identifier for a certain customer assigned by ISP.
 \item[\ndt{Node\_Managment\_ID}] Identifier for NaDa $Node Management$, must be
   different from all \ndt{Customer\_IDs}.
 \item[\ndt{APP\_Slice\_ID}] Tuple (\ndt{Customer\_ID}, \ndt{APP\_ID})
 \item[\ndt{NaDa\_Resource\_ID}] Tuple (\ndt{Node\_Management\_ID}, nil) or \ndt{APP\_Slice\_ID}
 \item[\ndt{NaDa\_Register}] Registration Message sent from $NaDa Node$ to $NaDa
   Management$ after booting the node.
 \item[\ndt{NaDa\_Meta\_Data}] is used for the following purposes:
   
   \begin{itemize}
    \item
      Meta data describing the \ndt{NaDa\_Content} to be shared, the
      corresponding locations,  and 
       the tracker. Information whether locations reside in the ISP domain or in the
       NaDa network and  the fingerprint of the content must be part of the meta file and
       the meta file has to be signed by the ISP. 
    \item
       Request for measurement data collected by $Node Monitoring$
   \end{itemize}
 \item[\ndt{App\_Slice\_Policy}] defines access rights to $App Slices$ based on the
   \ndt{NaDa\_Resource\_ID}. Finer granularity of access rights has to be implemented
   by the customer if necessary.
 \item[\ndt{App\_Slice\_Configuration}] includes key for encyption of
  the  corresponding $Node Store$, and \ndt{App\_Slice\_Policy}.
 \item[\ndt{App\_Slice\_Data}] Data stored in $Trusted Data Store$ assigned to a
   certain $App Slice$ (\ndt{App\_Slice\_Configuration}). It also includes monitoring data. Any access to such data will require to unseal the $Trusted Data Store$.
 \item[\ndt{NaDa\_Configure}] Command sent from $NaDa Management$ to $Node Management$
   to configure $App Slices$. It requires following parameters:
   \begin{itemize}  
     \item \ndt{App\_Slice\_Policy}
     \item Fingerprint of $App Slice$
     \item Command: activate / deactivate / restart $App Slice$
   \end{itemize} 
 \item[\ndt{NaDa\_Content}] describes both $App Slice$ to be installed, with the corresponding policy
   as well as control data of P2P protocol.
 \item[\ndt{App\_Content}] Content distributed by the customer and control data of
   the customer protocols.
 \item[\ndt{App\_User\_Request}] Command sent from $User$ to $App Slice$. The format of
   this data has to be described in the NaDa user interface specification.
 \item[\ndt{App\_User\_Response}] Response sent from $App Slice$ to $UI$-User Interface (a module of the $Node Management$). The format of
   this data has to be described in the NaDa user interface specification.
 \item \ndp{NaDa\_Log} (time stamp, action / measurement) produces a log entries for certain actions or measuring data. Actions and measurements are digitally signed by $Node Management$ relying on trusted time stamps.
  \item \ndp{Log\_Request} Command send from $App Slice$ to $Node Management$
     to request (Monitoring) measurement data. \ndp{NaDa\_Log} is used when creating the request message.
 \item \ndp{Log\_Response} Response sent from $Node Management$ to 
      $App Slice$ to forward information (e.g. Monitoring data) to $App
      Slice$. \ndp{NaDa\_Log} is used when creating the response message.

\end{description}

\begin{figure}[H]
    \centering
    \includegraphics[scale=1.0]{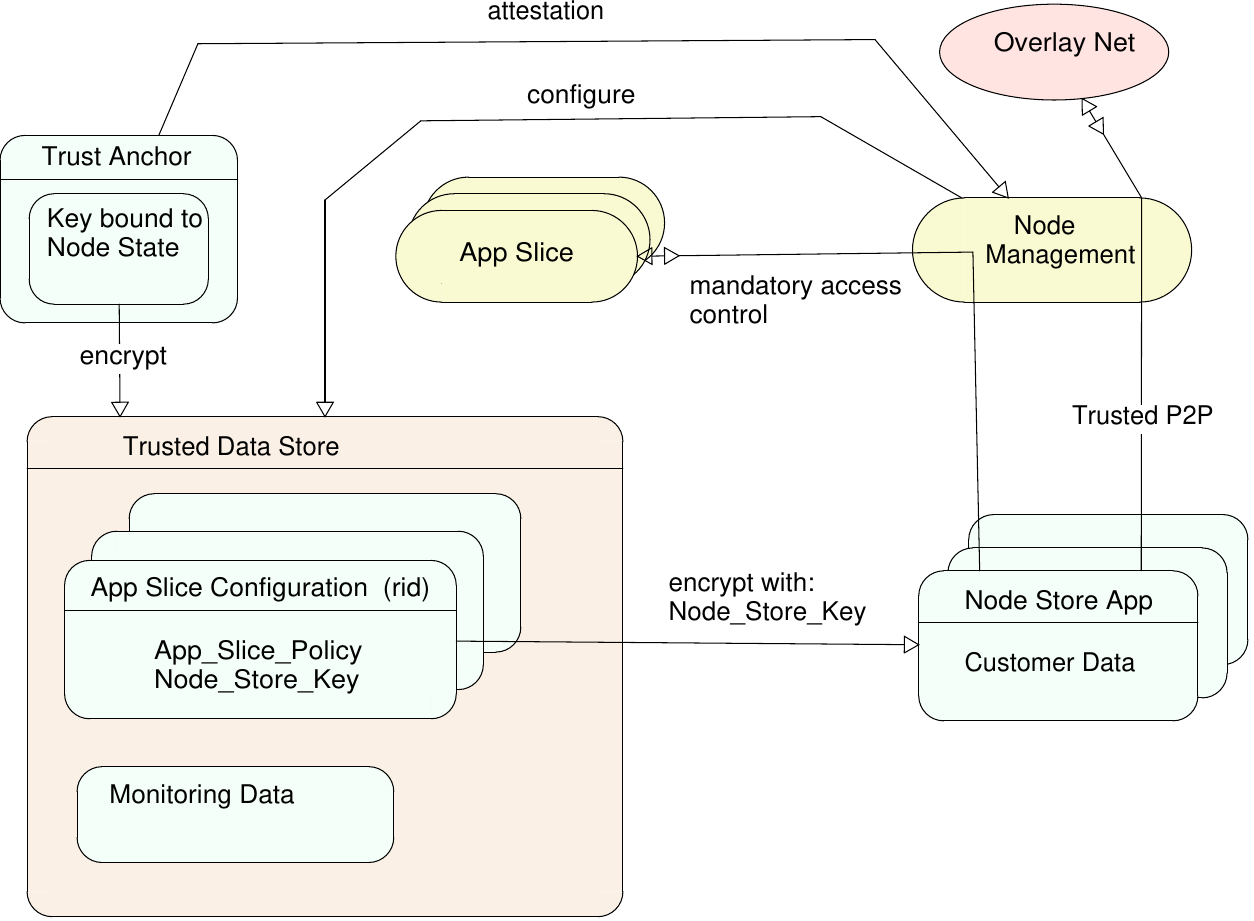}
    \caption{Overview Node Security Architecture }
    \label{fig:SecOverview}
\end{figure}

{\bf NaDa Primitives} describe security relevant actions executed by NaDa Nodes,
especially those relevant for communication between entities in different trust
boundaries. The primitives will process certain protocols, and compute or access
security relevant data. Figure \ref{fig:SecOverview} gives an overview of the
security mechanisms realized by the these primitives. All stored data will be
encrypted. Configuration of the NaDa Node and access of $App Slices$ to
resources is controlled by the $Node Managment$. Communication between $NaDa
Nodes$ is performed through overlay nets of the individual customers. Authentication
and computation of keys for encryption in the overlay net is realized based on trusted P2P protocol presented in \cite{kuntze:fuchs:rudolph:2010}.

\begin{itemize}
   \item \ndp{NaDa\_Authentication} ($NaDa Management$) Realizes
     the authentication and attestation described in \cite{kuntze:fuchs:rudolph:2010} and provides a symmetric key for the
     encryption of communication between the participating entities.
     \ndp{NaDa\_Authentication} is also executed after reception of 
     \ndp{NaDa\_Meta\_Data} e.g. for establish a encrypted connection 
     to respond requests for measurement data.
   \item \ndp{NaDa\_Get\_Ticket}  ($Node Management$, 2nd
     node,\ndt{NaDa\_Resource\_ID})
     realizes the the delivery of the trusted ticket (step 1 in MSC Figure 4 of 
     the trusted P2P protocol) to $Node Management$  which can be used for
     communication with  a second NaDa Node. 
   \item \ndp{Node\_Authentication}  (2nd node) Ticket obtained by
     \ndp{NaDa\_Get\_Ticket} is used for authentication to second node. After
     attestation of the second node a symmetric key for the encryption of the 
     communication between the two
     nodes is computed.  ``resource'' used in MSC Figure 4 of  the trusted P2P 
     protocol addresses the corresponding overlay net. The policy
     (\ndt{App\_Slice\_Policy}) for access to the overlay net has to be checked. 
   \item \ndp{NaDa\_Connect}(2nd entity,\ndt{NaDa\_Resource\_ID}) provides a symmetric key for
     encryption of communication. Depending of the type of the 2nd entity 
     only \ndp{NaDa\_Authentication} can be used if the 2nd entity resides in the the 
     domain of the ISP. If another $NaDa Node$ is addressed a trusted ticket
     (\ndp{Node\_Authentication}, \ndp{NaDa\_Get\_Ticket}, \ndp{Node\_Authentication}) has to
     be used. The information in which domain the entity resides is part of the
     NaDa\_Meta\_Data.
   \item \ndp{NaDa\_Register} ($NaDa Managment$) During the boot
     process $Node Management$ sends a registration message to to $NaDa
     Management$ after authentication. $NaDa Management$ responds with
     \ndt{NaDa\_Configure} to provide the node with the latest policy information.
   \item \ndp{NaDa\_Compute\_Key} (\ndt{NaDa\_Resource\_ID}) computes key for storage
     encryption bound to current  platform state and to the $App Slice$
     determined by \ndt{NaDa\_Resource\_ID}. 
   \item \ndp{NaDa\_Get\_Key}  (\ndt{NaDa\_Resource\_ID})
      Get key for storage encryption bound to a certain $App Slice$ defined by 
     \ndt{NaDa\_Resource\_ID}.
   \item \ndp{Nada\_Start\_P2P}(\ndt{NaDa\_Meta\_Data},\ndt{NaDa\_Resource\_ID}) starts the P2P
     protocol to download \ndt{NaDa\_Content} described  by \ndt{NaDa\_Meta\_Data} using the
     overlay net described by \ndt{NaDa\_Resource\_ID}. For every communication
     \ndp{NaDa\_Connect} has to be used to access the overlay net. 
\end{itemize}

\begin{figure}[H]
    \centering
    \includegraphics[scale=1.0]{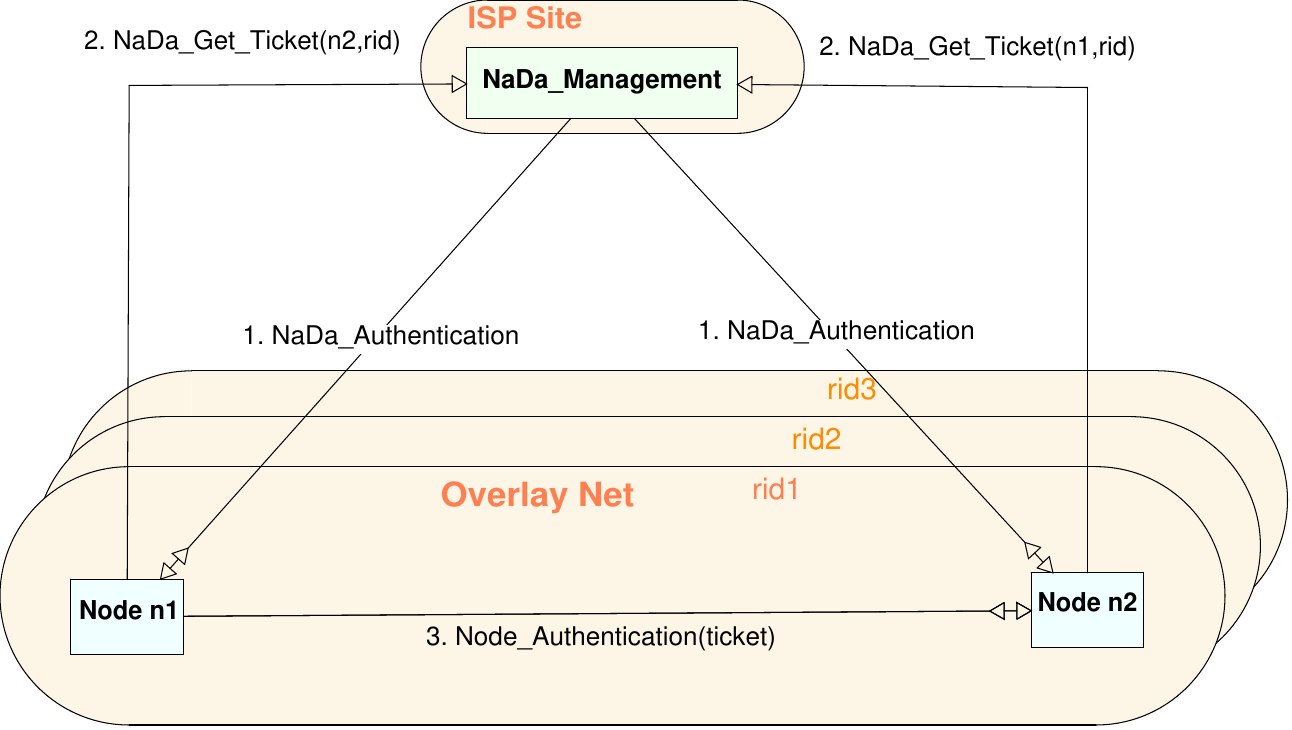}
    \caption{Overlay Net Constitution}
    \label{fig:Overlay Net}
\end{figure}

Figure \ref{fig:Overlay Net} explains the usage of the primitives to realize the
overlay net used for communication between NaDa Nodes in the case of a
centralized \ndt{NaDa\_Managment}. "rid" denotes the \ndt{NaDa\_Resource\_ID}
labeling the overlay net. To hide the structure of the net \ndt{NaDa\_Nodes} use
the primitive \ndt{NaDa\_Connect} (which is not depicted in this diagramm) to
establish a connection either with \ndt{NaDa\_Authentication} or with
\ndt{Node\_Authentication}. \ndt{Node\_Authentication} uses the primitives
\ndt{NaDa\_Authentication} and \\ \ndt{NaDa\_Get\_Key} for getting the ticket to
establish a connection to a second node.

\subsection{Putting Node into Service}
The following steps describe the process of setting up a (new) STB into a NaDa platform.

\begin{enumerate}
  \item Node reboot 
  \item Mutual authentication (\ndp{NaDa\_Authentication}) $NaDa Management \leftrightarrow Node Management$
  \item \ndp{NaDa\_Register} $Node Management \rightarrow NaDa Management$
  \item In certain time intervals: \ndp{NaDa\_Log} (\ndt{NaDa\_Measure\_Data})  $Node Monitoring
    \rightarrow NaDa Management$ to provide incentive mechanisms.
  \item For all $AppSlices$ $NaDa Management$ performs the following actions.
    \begin{itemize}
    \item Read \ndt{App\_Slice\_Policy}(\ndt{NaDa\_Resource\_ID}) from $Trusted Data Store$
     \item Disable traffic between new slice and all other slices not
           authorized by \ndt{App\_Slice\_Policy}.
    \item Configure overlay network according to the policy.
    \item Assign encrypted block device $Node\_Store$(\ndt{NaDa\_Resource\_ID}) to $App
      Slice$. The key from the corresponding \ndt{App\_Slice\_Policy} is used
      for encryption.
    \item Boot $App Slice$ image.
    \end{itemize}
\end{enumerate}

\subsection{App Slice Installation}

The installation of a customer $App Slices$ is performed in the following steps:

\begin{enumerate}
  \item \ndt{NaDa\_Meta\_Data} $NaDa Management \rightarrow Node Management$
  \item $Node Management$: \\
     \ndp{Nada\_Start\_P2P}(\ndt{NaDa\_Meta\_Data},\ndt{NaDa\_Management\_ID}). The following
     primitives are used between entities which are used in the DFDs for STRIDE analysis:
     \begin{itemize}
       \item  \ndp{NaDa\_Authentication} $Node Management \leftrightarrow NaDa
         Management$
      \item  \ndp{NaDa\_Get\_Ticket} $Node Management \leftrightarrow NaDa Management$
      \item  \ndp{Node\_Authentication} $Node Management \leftrightarrow Node Management$
      \item  \ndt{NaDa\_Content} $Node Management \leftrightarrow NaDa Management$
      \item  \ndt{NaDa\_Content} $Node Management \leftrightarrow Node Management$
     \end{itemize}
  \item After  the slice is downloaded and the fingerprint ist checked $Node
    Management$ installs the slice:
     \begin{itemize}
         \item Install Xen image
         \item Create \ndt{App\_Slice\_Configuration} using policy delivered with 
            $App Slice$.
         \item Disable traffic between new slice and all other slices not
           authorized by policy delivered with $App Slice$.
         \item Configure overlay network according to the policy.
         \item Create key (bound to current state) for encryption of virtual
           block device ($Node Store$) connected with the $App Slice$ and store
           key in the corresponding   $App Slice Configuration$.
         \item Store  $App Slice Configuration$ in $Trusted Data Store$ tagged
           by \ndt{NaDa\_Resource\_ID}.
         \item Assign block device ($Node Store$) to $App Slice$.
         \item Boot slice.
      \end{itemize}
   \item \ndt{App\_Log}(App Slice fingerprint and policy) $Node Management \rightarrow App Slice$
\end{enumerate}

\subsection{User Requests - Content Download}
This scenario considers an (end) user requesting, through an user interface, access to a multimedia content stored on a NaDa STB which in turn may be located in her domicile.

\begin{enumerate}
  \item \ndt{APP\_User\_Request} (for content) $User \rightarrow UI (Node Management)$
  \item \ndt{APP\_User\_Request} $UI (Node Management) \rightarrow App Slice$
  \item \ndt{NaDa\_Log}(\ndt{APP\_User\_Request},time stamp) $Node Management \rightarrow App Slice$:
    (signed by ISP)
  \item $App Slice$ start P2P download (overlay net configured by $Node
    Management$ and $NaDa Management$ otherwise transparent for $Node Management$)
  \item \ndt{APP\_User\_Response} ($App Slice \rightarrow UI (NaDa Management)$, if
    content, or certain part of content is downloaded)
  \item \ndt{APP\_User\_Request} $User \rightarrow UI (Node Management)$:  (e.g. Play)
  \item \ndt{APP\_User\_Request} $UI (Node Management) \rightarrow App Slice$:  (e.g. Play)
  \item \ndt{NaDa\_Log}(\ndt{APP\_User\_Request},time stamp)  $Node Management \rightarrow App Slice$:
    (signed by ISP)
  \item $Node Management \rightarrow App Slice$ (in defined time intervals): \ndt{NaDa\_Log}(play,time stamp) signed
    by ISP
  \item \ndt{APP\_User\_Request} $User \rightarrow UI (Node Management)$:  (e.g. Stop)
  \item \ndt{APP\_User\_Request} $UI (Node Management) \rightarrow App Slice$:  (e.g. Stop)
  \item \ndt{NaDa\_Log}(\ndt{APP\_User\_Request},time stamp)  $Node Management \rightarrow App Slice$:
    (signed by ISP)
\end{enumerate}

\subsection{Monitoring as Support for Operation and Management Tasks}
We consider the following three sub-scenarios in which the basic NaDa monitoring approach (see D1.1 \cite{D11}, section 3.3) is deployed for the purpose of monitoring operational and management tasks. The rationale behind this rests on the fact that, ISPs may want to monitor all NaDa related network traffic and resource access/consumption in order to improve (or at least permanently guarantee) the operational stability and reliability of its platform. In such a context, there are  interested in gathering STBs and slices related measurements (e.g. performance or anomalies) which are then made available to different stakeholders (e.g. content providers  or eventually end users). 

As examples of operation and management tasks, we consider the following three use-cases: Firstly, the low-cost resources (re-) allocation in the NaDa Platform, secondly the proactive and automatic detection of anomalies and load peaks, and finally the control of isolation between slices. 
These three perspectives on operation tasks pinpoint the relevance of monitoring as process which, on one hand provides inputs for the analysis and the improvement of the quality assurance (failure detection, maintenance, performance tuning) especially in video-on-demand context, and on the other hand helps reducing the network access cost.

\subsubsection{Low-cost resources (re-) allocation} 
A persistent and global view of resources consumption and availability is fundamental for performing a flexible resources (e.g. available free slices, bandwidth or storage capability) allocation to content providers (i.e. their respective slices) in the NaDa platform. The allocation task in turn mainly relies on information collected by the NaDa monitoring processes. Here, a CP that operates slices in geographically distributed nodes, for instance across Europe is interested to know the state of health (e.g. running applications, security configuration and link utilization) of its nodes, characteristics of network paths between the nodes (e.g. reachability, delay, available bandwidth), and the current configuration/properties of its slices (e.g. memory, storage and computational utilization). Based on these information, the CP then identifies and analyzes the service popularity in a particular region (i.e. portion of the NaDa network). The gained knowledges allow CP to dynamically adapt its content distribution flow w.r.t end users preferences (e.g. based on service popularity), STB and slices configuration and actual SLA (which has been renegotiated in order for instance to allow CP to purchase more resources in the targeted NaDa network's portion). Based on the collected Information, the pre-loading strategies considered in NaDa is specified and implemented. 

\subsubsection{Proactive and automatic detection of anomalies and load peaks}
Beside the support of an efficient resources (re-) allocation, the observation of both network anomalies (e.g. indication of DDOS, failure of STB) and workload peaks is highly important for security, platform and resource management purposes. In this context, the required information are collected platform-wide, for instance (i) in a portion of the NaDa network or (ii) inside a hardware module (e.g. STB, residential gateway, DSL-Line Access Multiplexer, IP backbone). Here, a near real time notification and reporting of failure provides a significant level of flexibility, e.g. it allows the CP to automatically adapt to the new situation through the reconfiguration or isolation of both hardware and slices. The ISP is therefore interested to have both a historical and statistical view of platform's behavior (inclusive anomalies, CP related resources consumption or even illicit attempts to obtain more resources). 

\subsubsection{Isolation management}
Since one central aspect of NaDa is to provide different CPs with the ability to carry their respective application slices on the same NaDa node, it is crucial from their point of view to rely on strict but flexible isolation of rivals applications. Considering the case in which for instance Warner Bros Entertainment\footnote{\url{http://www.warnerbros.com/}} and Vivendi\footnote{\url{http://www.vivendi.com/}} have slices located on the same node. Here each provider is be allowed to access and collect any of their slices related properties (e.g. storage utilization). In absence of suitable, access control mechanisms and well-defined isolation between the different Monitoring flows, Warner Bros Entertainment may gain access to monitoring logs in which for instance the link utilization of the common host node is reported. This information does not only provide a view on the common node, at a given time, but also details for instance about services offered by Vivendi. Through an analysis of link utilization data, Warner Bros Entertainment can then gain access to business strategic information like statistics about services type and popularity, or the geographical region of interest of services offered by its competitor, Vivendi. Additionally monitoring processes furnish both NaDa and Node Management required inputs about a STB's state,  allocated resources and workload of each application slices. Based on such information, the management modules then enforces isolation between slices, i.e., prohibition of non authorized access to sensitive competitors' data or to access platform's resources.

\section{Security Requirements and assumptions}
%TODO introduction to assumptions and requirements
The NaDa security architecture is based on several security requirements 
derived from the use cases. In the design of the security architecture
different assumptions on security functionalities are done. In the following
sections security requirements and assumptions relevant for the architecture
are presented.

\subsection{Security Requirements}\label{sec:Security Requirements}
The following high level security requirements for the NaDa architecture form
the determining factor for the identification of threats and the appropriate
mitigation techniques when the STRIDE analysis is conducted:

\begin{itemize}
\item Isolation of App Slices corresponding to the policies defined by the
   customers must be ensured. Mandatory access control to network and to customer
   content has to be enforced according to these policies (Authentication,
   Authorization, Virtualization).

\item Customer content must be protected against manipulation and illegal access
   during network transfer and storage on NaDa nodes (Data Integrity, Confidentiality).

\item Customers must be enabled to set up secure accounting (Non-repudiation services).

\item Nodes participating in the NaDa network must perform attestation, to
   report the integrity of the software running on these nodes (Platform Integrity).

\item Customer software should only be activated after checking the corresponding
  ISP certificate (Authenticity).
  
\item Correctness and Integrity of Monitoring Data: Requestor should have the confidence that the monitoring data reported are indeed what was monitored. The accuracy of the information reported has to be guarantee, since inaccurate data may lead to bad analysis results. Such guarantee may rely on one hand on the monitor's ability to measure and report accurate data and on the other hand on the controller's ability to verify authenticity and freshness of reported data. It may be assumed that unauthentic or unnecessary data are filtered by the controller before storage in the central MIB. 

\item Confidentiality and Fine-grained Access Control: As NaDa considers outputs of monitoring processes as vital inputs for operational and management tasks (i.e. real-time performance analysis and troubleshooting) and such inputs as highly sensitive from the point of view of ISPs and CPs, confidentiality and access control methods have to be integrated within all phases of the monitoring process. The confidentiality has to be ensured from the early stage of the data collection up to the reporting and disclosure phases. The Collector orchestrating the data collection has to rely on encryption primitive  or anonymization technique in order to allow a confidential reporting of sensitive monitoring data.  The anonimization or encryption of such data has to be performed without loss of effectiveness and without a deterioration of quality and accuracy of collected data. A de-anonymisation or deencryption of stored data should only be possible for components and processes with appropriate (cryptographic) credentials. On the other hand, access control methods may rely on attributes (NaDa resource identifier or role), platform integrity or slice's level of trust, ensuring that only legitimate and authorized entities can access the collected measurements.

\end{itemize}

Since the nodes are operated in different environments different security
requirements are given for these nodes from the point of view of the ISP and of
the Customer.  

From the point of view of the ISP it's very important to avoid
product recall or cost-intensive service for NaDa Nodes. Also the ISP must give
guaranty that App Slices of different customers are strictly isolated. Thus it's
in the interest of the ISP to control installed App Slices and their updates.
If for instance exploits enabling elevation of privileges for virtual images
would be emerged it should not be possible for customers to install software
using these exploits. Thus only App Slices signed by the ISP should be installed
on NaDa Nodes in the field.  

The situation is somewhat different for NaDa Nodes under control of the
customer. Trust boundaries between App Slices of different customers running on
the same node do not exist in this case.  Injection of content must be possible
for these nodes. The main requirement is to restrict the access of the
corresponding App Slices to the corresponding overlay net defined by the
resource identifier of the APP Slices. Thus the integrity of the resource
identifier of the corresponding App Slices has to be assured to prevent breaking
out of the overlay net. To ensure this requirement the customer software of
these specialized nodes also has to be certified by the ISP. Other solutions
where uncertified hardware and software is utilized would require control
possibilities for access to the overlay net of these entities. Various
solutions, also depending from the contract between ISP and customer, would be
possible but are out of scope of this analysis.

The permission to install updates on App Slices for customers would exclude
the possibility for the ISP to check the integrity of the App Slice software installed
on the nodes. Thus arbitrary manipulation of App Slice software by users would
be alleviated.

A detailed discussion on the security requirements relevant for the NaDa 
scenarios can be found in D1.1 \cite{D11} \cite{kuntze:repp:2009} \cite{Fuchs:Guergens:Rudolph:2009:pattern} \cite{Fuchs:Guergens:Rudolph:2009} and \cite{kuntze:fuchs:rudolph:2009}.

\subsection{Security Assumptions}\label{sec:Assumptions}
For the scenarios presented in section \ref{sec:Use Scenarios} following assumptions apply:
\begin{itemize}
  \item Sensitive data, like private ISP keys, App Slice fingerprints, and App slice
    images to be delivered will be protected by the ISP. The corresponding
    measures will not be part of this threat analysis.
  \item Manipulation of the software e.g.\ due to a buffer overflow can't be
    prevented. 
  \item Customer IDs and App Slice IDs  must be part of the tickets computed by
    the trusted peer to peer protocol, to provide  addressing of Node Management
    and App Slices and also for
    the configuration of the overlay network. This address information will be
    stored in ``$NaDa\_Resource\_ID$'' used in MSC Figure 4 (Step 1, as
    ``resource'') of the trusted peer to
    peer protocol defined in \cite{kuntze:fuchs:rudolph:2010}. NaDa\_Resource is defined as follows: 
    \begin{tabbing}
    $NaDa\_Resource \ \  \  \ \ \  \  \ $ \= $:= (Node\_Management\_ID,nil) \ | \ APP\_Slice\_ID$\\     
    $APP\_Slice\_ID$ \= $:= (Customer\_ID,APP\_ID)$
    \end{tabbing}
  \item During boot process of a NaDa Node the internet connection to the ISP must be
    available, to register the node and for time synchronization.    
  
\item Trusted Platform Administrator: The designed isolation administrator and
NaDa platform manager is trusted to access and   process sensible platform and
applications' metering data. Those managers act in accordance with their
costumers' security preferences.
  
\item Correct hardware: The underlying hardware (e.g. TPM, CPU-chip, I/O and
storage devices, etc.) behaves in accordance with the specifications and standards.
	
\item Trusted TCB and reliable Isolation: The virtualization engines as well as
other components of the trusted computing base of the NaDa monitoring architecture are correct and behave as specified.

\item No advanced physical attacks: We have assumed that the underlying hardware included in the NaDa Monitoring Architecture are resilient against such type of attacks. \\

\end{itemize}

\section{High level model}
The high level model for the NaDa architecture is composed of the node 
architecture including the communication of the interactions between the 
respective components on the one hand and the security model for the monitoring
modules. The monitoring modules are situated and operated based on the node architecture
as an independent service.  
In the following chapters these two models are presented.

\subsection{Node architecture model and interactions}

\begin{figure}
\centering
\includegraphics[width=0.75\textwidth]{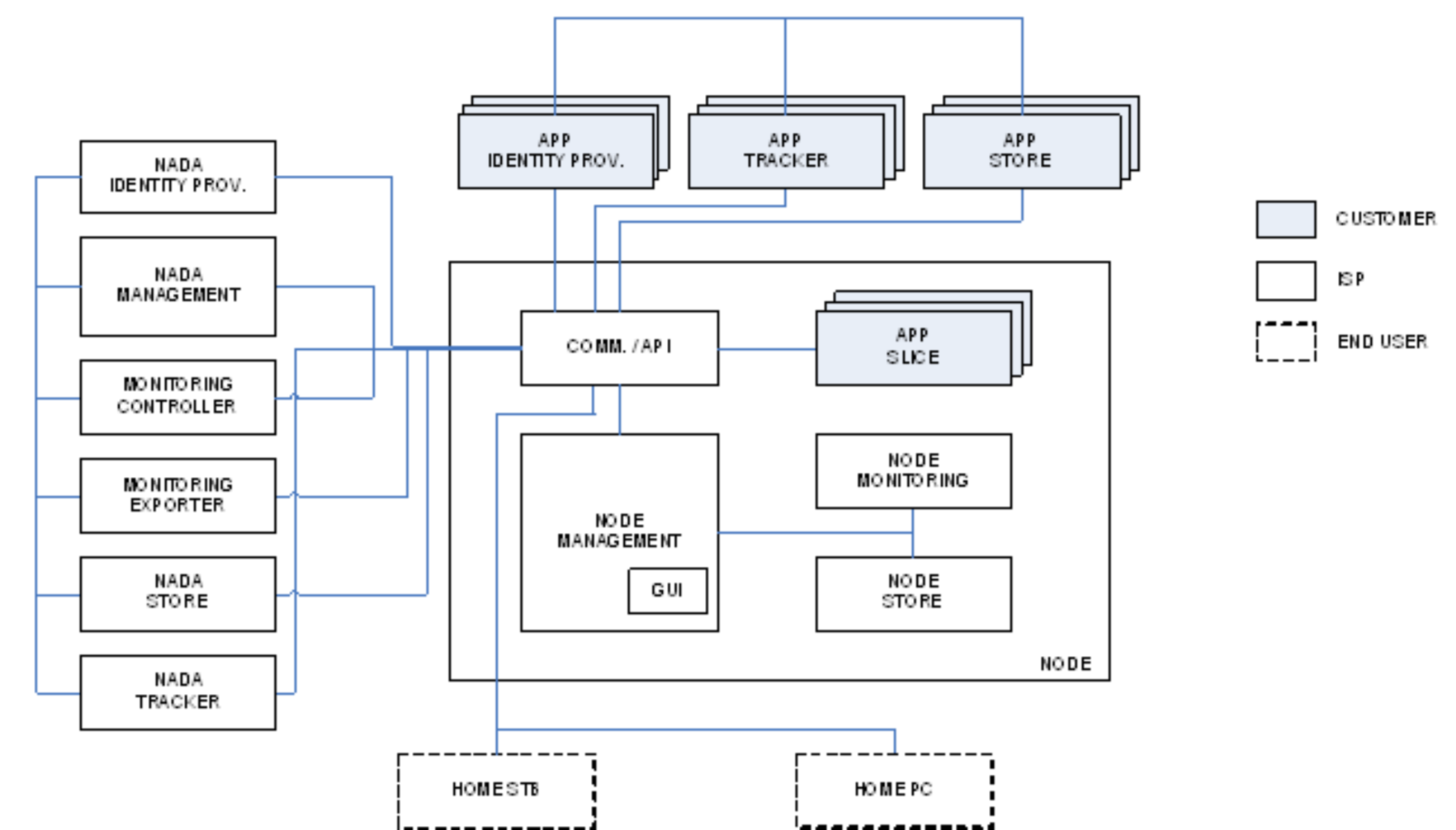}
\caption{High level architecture for the distribution of content}
\label{fig:high_level_architecture}
\end{figure}

As presented in \cite{kuntze:repp:2009}\cite{D11}\cite{D31} the 
architecture of the NaDa infrastructure is based on the P2P paradigm
allowing for a highly available infrastructure by distributing some of
the core functionalities into the nodes installed. The resulting high
level architecture is depicted in Figure~\ref{fig:high_level_architecture} and consists of the node and
its supporting infrastructure. Mainly two players{\textquoteright}
components are shown there. The customer is represented by its slice
and the corresponding centralized application tracker as well as its
support by the customers Identity Provider and the store of content
offered by the customer. In Figure \ref{fig:high_level_architecture} 
only one customer is shown for the
sake of simplicity. Additional customers are duplicating the customer
related components.

\subsection{Security Mechanisms for the NaDa Monitoring Architecture} %Modules
This section describes a secure monitoring infrastructure using trusted computing and virtualization concepts. Both concepts enable trustworthy generation and storage of monitoring data. The further disclosure of such sensitive data is governed in the proposed architecture by fine grained access control policies. In order to provide the fine grained access control required for a large scale use of NaDa, we have integrated XACML \cite{xacml02} components into the NaDa basic monitoring architecture. The new security mechanisms extend the architectural view on the basic NaDa Monitoring Architecture (see D1.1 \cite{D11}\cite{D31} and \cite{kuntze:repp:2009}). %\ref{kuntze:repp:2009}
\begin{figure}
\centering
\includegraphics[width=1.0\textwidth]{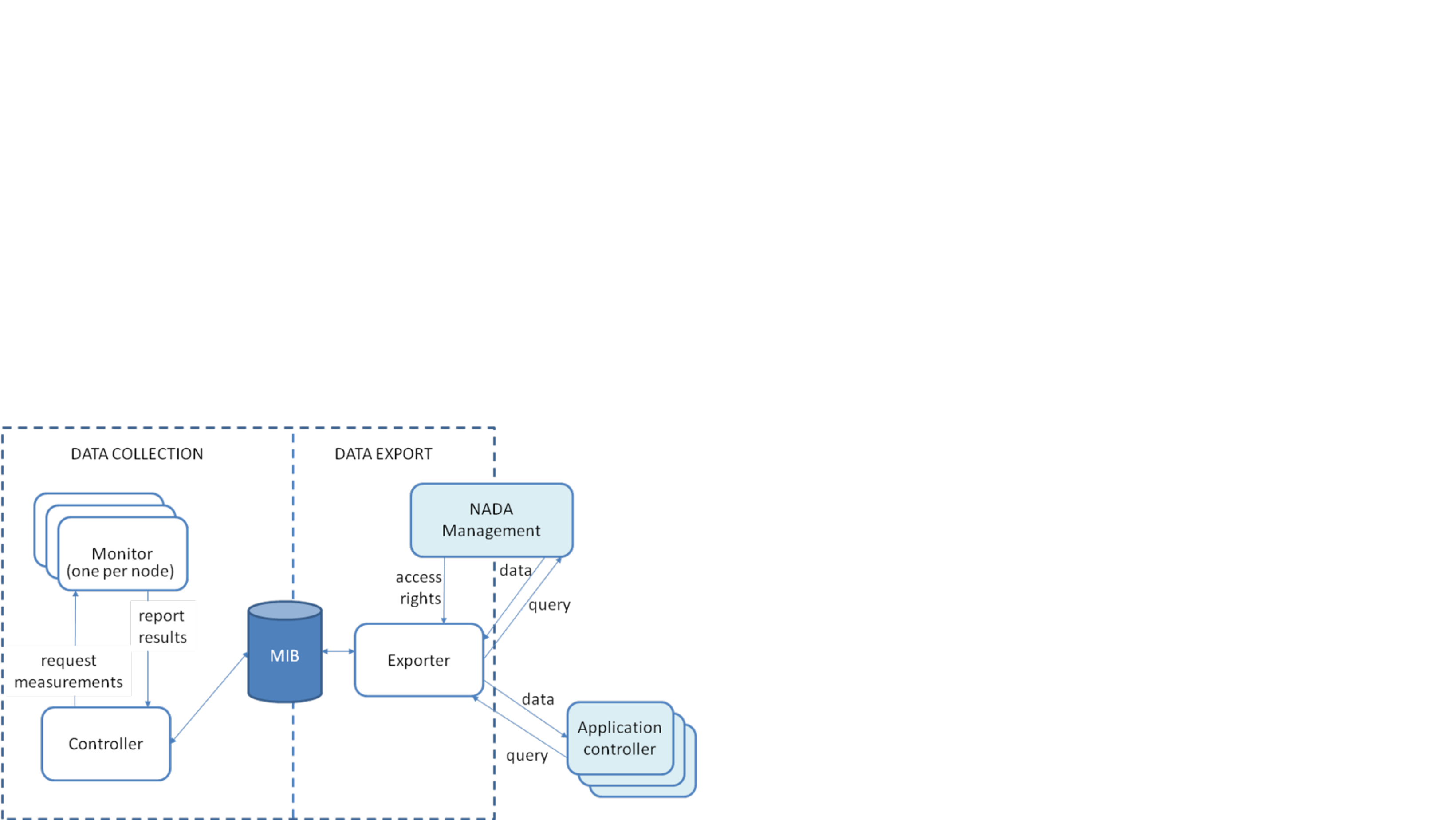}
\caption{High level architecture for the measurement}
\label{fig:high_level_measurement}
\end{figure}

\subsubsection{Internal Security and Access Control Mechanisms}

The proposed mechanisms ensuring STB internals data protection leverage the isolation techniques already introduced (see D1.1 \cite{D11}\cite{D31}) and allow fine grained access control to monitoring measurements. Such measurements generated during the active monitoring of a STB as well as during the passive monitoring of slices running in that STB are temporarily
stored in a local repository which in turn is protected by the means of secure storage (NADA security
primitive \emph{NaDa\_Compute\_Key} and \emph{NaDa\_Get\_Key} as defined in section \ref{sec:DFDs}).
Previously to their storage the monitoring data are signed by the Node
Management relying on trusted time stamps as specified by the \emph{NaDa\_Log} primitive
\ref{sec:DFDs}. Any request from a slice for internal
measurements is handled by the Node Management, which is internally responsible
to authenticate request for local measurements and to restrict access to locally stored monitoring
data. The request and access to monitoring data are performed with respect to
authentication, authorization (section \ref{sec:internalsProcess}) and confidential disclosure of monitoring data. 

\begin{figure}
\centering
\includegraphics[width=0.75\textwidth]{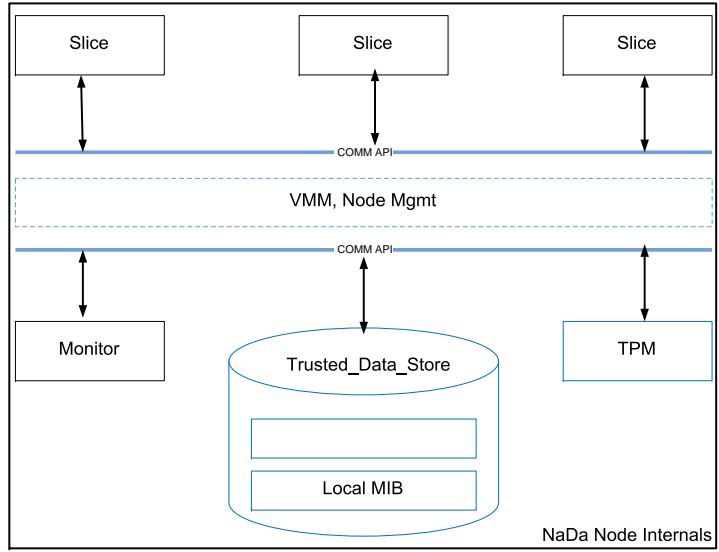}
\caption{ Node Intermal Security Components}
\label{fig:internals}
\end{figure}

Figure \ref{fig:internals} depicts the relevant security components and modules
of the STB focusing on the protection of measurements locally stored.

In the following a description of those components and modules as well as their
respective contributions to secure the internal monitoring process is provided.
The first two modules (monitor and local MIB) basically extents the functions of
the Node Monitoring initially introduced in D1.1 \cite{D11}\cite{D31} section 3.1.
\paragraph{Monitor}

The monitor deployed in each NaDa node (STB) is responsible for triggering active
measurements from this node and for passively monitoring slices running in the
node. It captures several types of node and slices related information, protects
them according to (NaDa and slices) policies and delivers them to monitoring
service. Relying on traditional tools (e.g. CACTI \cite{cacti}, Nagios \cite{nagios} or Munin \cite{munin}) which in turn
use standards like PR-SCTP or IPFIX, the monitor initiates the collection of host
node, network characteristics as well as slices related metering data.
\paragraph{Local MIB}

All STB's and local slices' measured properties are first stored in a local MIB.
The content of each local MIB is then transferred in the global MIB. The split
between local and global MIB as well as the periodical update of the gloabl MIB can be considered as a step  towards a good compromise between the need
for fresh platform measurements and the monitoring overhead.  The secure storage
by means of integrity and confidentiality relies on primitive for storage
encryption bound to the STB' configuration or to slice's state. This process
benefits from the combined potentials of TPM and virtualization as specified in
D1.1 \cite{D11}\cite{D31}.
\paragraph{VMM}

VMM primarily provides an abstraction to the underlying hardware resources of the
host STB, and defines an isolated execution environment for each stored slices.
In order to perform the isolation between application slices running on the same
physical NaDa box, the VMM defines a module which is build as policy enforcement
and policy decision point for a wide range of low-level security policies. For instance
this module may include pre-defined policies that specify which slices are
allowed to communicate or share resources together and which security
requirements the exchanged messages must satisfy. The specification and enforcement of high-level
(finer-grained) access control requirements for the temporarily stored local measurements will be handled by the node monitoring.
\paragraph{TPM}

We refer to the TPM as hardware-based trust anchor inside the STB (see D1.1 \cite{D11}\cite{D31}). The overall security mechanisms for the NaDa Monitoring make use of its potential for several purposes e.g., for a trustworthy collection of boxes and
slices properties, and for ensuring a continuous chain of trust/integrity up to the slices, and 

\subsubsection{Internal Monitoring Process}
\label{sec:internalsProcess}
The workflow for requesting internal monitoring data (e.g. performance of host box or characteristics of co-located slices) is depicted in Figure \ref{fig:internalsProcess} and described as follows:

\begin{itemize}
  \item First of all, a slice S1 requests measurement data (\emph{Log\_Request} and \emph{Log\_Response} in \ref{sec:DFDs}). Upon receiving this request, the Node Management applying the isolation policy also performs internal access control to the local MIB. This involves evaluating the slice's request by means of authentication (cf. relying on slice's unique identifier \emph{APP\_Slice\_ID}). As result of this step, Node Management which is designed as Policy Enforcement Point (PEP) validates S1's level of trust and allows the request to be forwarded (asking for an authorization decision) to the Node Monitoring (steps 1-3).   
\begin{figure}
\centering
\includegraphics[width=1.0\textwidth]{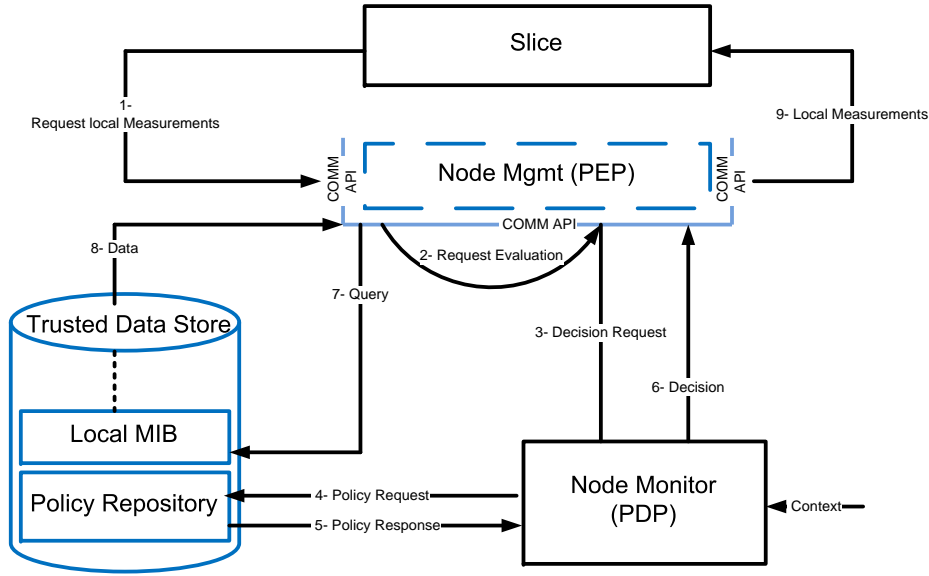}
\caption{Internals Monitoring Process}
\label{fig:internalsProcess}
\end{figure} 
  
  \item Node Monitoring designed as Policy Decision Point (PDP) then performs access authorization providing the required finer granularity of access control. This involves requesting policy rules and obligations from the policy repository.  Based on these policies and other security context (e.g. trust level of slice or current state of the STB) the Node Monitoring generates an authorization (permit or deny) decision which is then transferred to the Node Management(PEP) (steps 4-6). 
 
  \item Finally, the PDP's decision is enforced by the PEP (i.e. Node Management) which unseals (relying on \emph{NaDa\_Get\_Key}) the local MIB from the secure storage, queries and transfers the targeted measurements information to S1 (steps 7-9).
\end{itemize}

\subsubsection{External Security Mechanisms}
\begin{figure}
\centering
\includegraphics[width=0.75\textwidth]{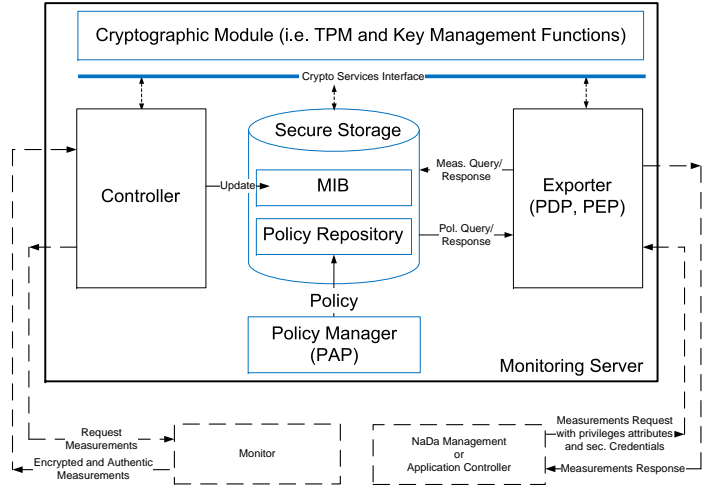}
\caption{Extended NaDa Monitoring Service Architecture}
\label{fig:MonitServer}
\end{figure}

As shown in Figure \ref{fig:MonitServer}, this section considers mechanisms to enforce security requirements of a central component of the NaDa Monitoring Architecture: The monitoring server. The resulting security methods aim at providing the primitive to protect the integrity and confidentiality of the monitoring data from the reporting by STBs up to their storage in the global MIB and their dissemination by the Monitoring Server. As result we extend the functionalities of monitoring server's main component as defined in the following paragraphs.

\paragraph{Secure Reporting}
In our extension, Node Monitoring reports the monitoring measurements as encrypted data. The secure reporting by means of encryption and digital signature mainly relies on \emph{NaDa\_Connect} primitive for building the encrypted communication channel between monitoring in STB and Controller in monitoring server. The controller module of the monitoring server on the other hand decrypts the received measurement values, performs the orchestration and stores the plain-text data in the MIB. Previously to the encryption and decryption, both entities performed a bidirectional authentication (using the \emph{Node\_Authentication} primitive) as well as the negotiation of the symmetric key used to encrypt or decrypt  the reported measurement.  

\paragraph{Global Access Control Mechanisms}
The global access control mechanisms tailor the XACML Framework \cite{xacml02}to our purpose. The resulted architecture is depicted in Figure \ref{fig:MonitServer} and mainly includes four components: Controller, MIB, Exporter which is design as PEP and PDP, and a Policy Manager which is designed to be in charge of the security policies governing the access to the MIB. 

Although Figure \ref{fig:MonitServer} depicts a centralized access control approach describing the four components as part of a single central entity, we stress that these components can also be physically distributed throughout the NaDa network. The following subparagraphs detail the functionalities of each of these components.

\subparagraph{Controller}
Besides its ability to require measurements from different STB (i.e. monitors) and orchestrate all the measurement collection the Controller can also use en-/decryption primitives (\emph{NaDa\_Compute\_Key} and \emph{NaDa\_Get\_Key}) to store the collected measurements as clear text in the global MIB. The required guarantee that the monitoring server interacts with the correct STB (i.e. the box is trustworthy and belong to the correct ISP domain) is provided when relying on \emph{NaDa\_Authentication} and \emph{Node\_Authentication} primitive (see \ref{sec:DFDs}).
\subparagraph{(Global) MIB}
This repository stores all STB' and Network measurements. These data are protected by means of encrypted storage which is bound to trust state and configuration of the monitoring server. We rely for this purpose on a set of functions specified by the TCG i.e., integrity reporting and attestation, (un-) sealing and (un-) binding functions among others (see D1.1 \cite{D11}\cite{D31} Section 2.3.2). Thus, the MIB can only be available if no change has been detected in the configuration of the monitoring server (i.e. the server and its measurement information data base have not been subject of security attacks, and the integrity of the stored monitoring data remains intact). 

\subparagraph{Exporter}
We extent the Exporter Module with a set of entities and functional components that allow authorization decisions to be made and enforced based on security credentials (e.g. certified requester's attributes) and in accordance to access control policies. In the following an overview of the PEP and PDP roles of the Exporter is presented.

\begin{itemize}
  \item Policy Enforcement Point (PEP) - The Exporter is the entity where the external query for monitoring data arrive. It authenticates the queries and enforces the decision made by PDP's decision. The PEP must be able to intercept any request, for monitoring data, between Application Controller or NaDa Management and the Exporter. It should be noted that, the technical implementation of the PEP must be performed such that the PEP cannot be bypassed in order to access the protected MIB.

 \item Policy Decision Point (PDP) - As PDP the Exporter makes decisions to authorize access. The PDP uses the access control policies from the PAP as well as additional information (context information, e.g. slice's trust level or server current configuration) in order to evaluates policies and make a decision to authorize access to the local MIB.
\end{itemize}

The authorization decision process also relies on other components like for instance the context handler, which are for simplicity reasons not shown in Figure \ref{fig:MonitServer}. As part of controlling access to the MIB, the Exporter also performs request evaluation/ authentication as already pointed out in the basic Monitoring architecture (see D1.1 \cite{D11}\cite{D31}). It therefore integrates the authentication function and primitives provided by the cryptographic module (see below).  

\subparagraph{Policy Manager}
This entity is designed as XACML Policy Administration Point (PAP) which is basically the entity that creates storages and manages all access control policies used by the PDP (i.e. Exporter). These policies consist of decision rules, conditions, and other constraints for accessing the data stored in the global MIB. The policy manager specifies access control policies with respect to the level of confidentiality required for monitoring information (neutral or sensitive), and security affiliation and credentials of the requester. Furthermore, the policy manager specifies conditions and obligations for access control policies (e.g. log any request and access to MIB for accountability purposes) allowing a deployment in a vast range of use-cases. This will provide the basis for future security auditing.

\subparagraph{Cryptographic Module}  
This module includes a TPM and other cryptographic and key management primitives. It therefore provides, validates, and maintains the cryptographic keys which are used by the data protection mechanisms enforced by the Exporter. For these purposes, the interaction with an external Public Key Infrastructure (PKI) is considered. In addition it provides along with the authentication primitive defined in section \ref{sec:DFDs}. the required inputs for trust negotiation between access requester and monitoring server, especially when the monitoring server needs slice to provide authentic credentials/attributes for authorization decision.

\subsubsection{External Monitoring Process}
\begin{figure}
\centering
\includegraphics[width=0.75\textwidth]{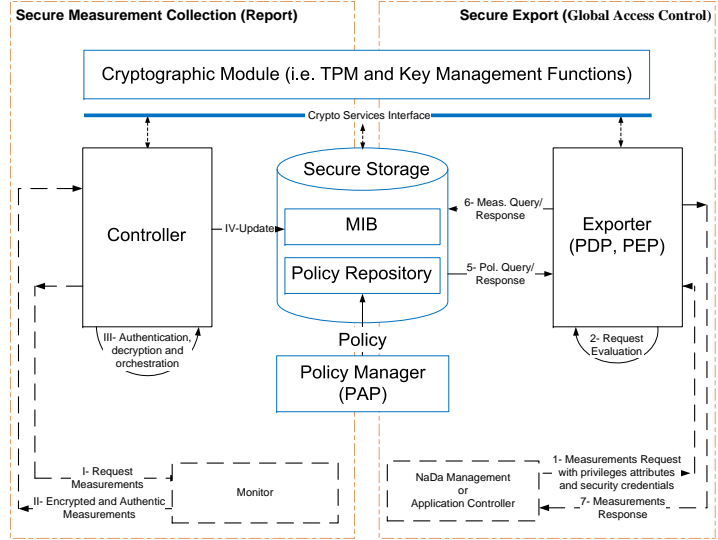}
\caption{External Monitoring Process}
\label{fig:ExternalsProcess}
\end{figure}

This process is relevant for Application Controller and NaDa Management which need information related to the performance/status of any NaDa box or related to (owned) slices stored somewhere on the NaDa platform. Figure \ref{fig:ExternalsProcess} shows the integration of security mechanisms described above and differentiate between two subprocesses: Secure measurement collection and secure measurement export.
\paragraph{Secure Measurement Collection (Report)}
The steps included in this subprocess are depicted in the left Block of Figure \ref{fig:ExternalsProcess}. 
\begin{itemize}
	\item In the first step, the Monitoring Server (through its Controller module) contacts the Monitor inside a STB and request measurements (\emph{Log\_Request})(step I).  

	\item In order to guarantee authenticity and confidentiality, both Controller and Monitor engage in a bidirectional authentication (\emph{Node\_Authentication}) and negotiate a symmetric key for encrypted communication (\emph{NaDa\_Connect}). The monitor then transmits STB's measurements (\emph{Log\_Response}) as ciphertext to the Collector (steps II, III).

	\item After receiving encrypted measurements, the Controller uses its part of the symmetric key to decrypt the ciphertext. Afterwards it orchestrates all data collection, unseals (\emph{NaDa\_Get\_Key}) and updates the MIB (steps IV, V).

\end{itemize}

\paragraph{Secure Measurement Export (Global Access Control)}
The right block of Figure \ref{fig:ExternalsProcess} shows the different steps which composed the Secure Measurement Export. 
\begin{itemize}
	\item At the beginning the NaDa Management or an instance of an Application Controller sends a measurement request (\emph{Log\_Request}) to the Exporter, along with its security credentials (step 1).

\item 	The Exporter funding as PEP evaluates the request by means of authentication (\emph{Node\_Authentication}) and retrieves authorization rules from unsealed Policy Repository (step 2, 3).

\item 	Based on these rules and addition information (e.g. current configuration) the Exporter designed as PDP makes a deny/allow decision. In case it authorizes the access, the targeted measurements are queried from unsealed MIB and transferred (\emph{Log\_Response}) to the NaDa Management/ Application Controller (step 4, 5).
\end{itemize}

\section{Threat Models}
This Deliverable (D3.2) defines the NaDa security architecture based on the STRIDE methodology given by
Microsoft. In the respective section STRIDE is introduced and the architecture
is given. As a mean to verify the design a second approach is introduced using
an abstract functional system model. Section \ref{sec:functionalSystemModel}
introduces this approach and provides first results later used in the evaluation
as part of D3.4 \cite{D34}.

\subsection{STRIDE methodology and architecture}
%TODO STRIDE introduction

{
A threat model of the NaDa architecture, based on STRIDE, a methodology
introduced by Howard and Lipner in \cite{Howard:Lipner:2006}, will be
presented. STRIDE is an acronym for Spoofing, Tampering, Repudiation,
Information Disclosure, Denial of Service, and Elevation of Privilege (EoP). \\
Two use scenarios build the base of the threat model. Data flow diagrams (DFDs)
will be created for the corresponding use cases. All primitives and the corresponding
elements of the NaDa architecture are included in the DFDs. For every element of
a DFD applicable threats will be assigned to these elements, according to the
stated high level security requirements for the NaDa architecture. In the next step the
threat mitigation techniques of the NaDa architecture will be assigned to the
DFD elements and their corresponding threats. Finally it can be checked, whether
there is a countermeasure for every expected threat assigned to the DFD. The
tables with the measurement assignment can be used as a checklist for the
implementation and code review of the security mechanisms of the NaDa components.}

\subsubsection{Differences to STRIDE Methodology}

Table \ref{tab:Mapping STRIDE Threats to DFD Element Types} shows the mapping of threats
to DFD elements  proposed in \cite{Howard:Lipner:2006}. This mapping was
inappropriate for several reasons:
\begin{itemize}
\item The abstraction level for processes required different measures against
  spoofing depending on the corresponding data flow entities. So spoofing was
also assigned to data flows.
\item Also repudiation was assigned to data flows in opposite to the standard way
  of proceeding, because all actions related to this threat could be mapped
  exactly to
  one data flow entity. Repudiation was not considered for external entities and
  processes. For data stores and processes tempering was checked for entities
  involved in the corresponding data flow.
\end{itemize}

\begin{figure}[H]
    \centering
    \includegraphics[scale=0.8]{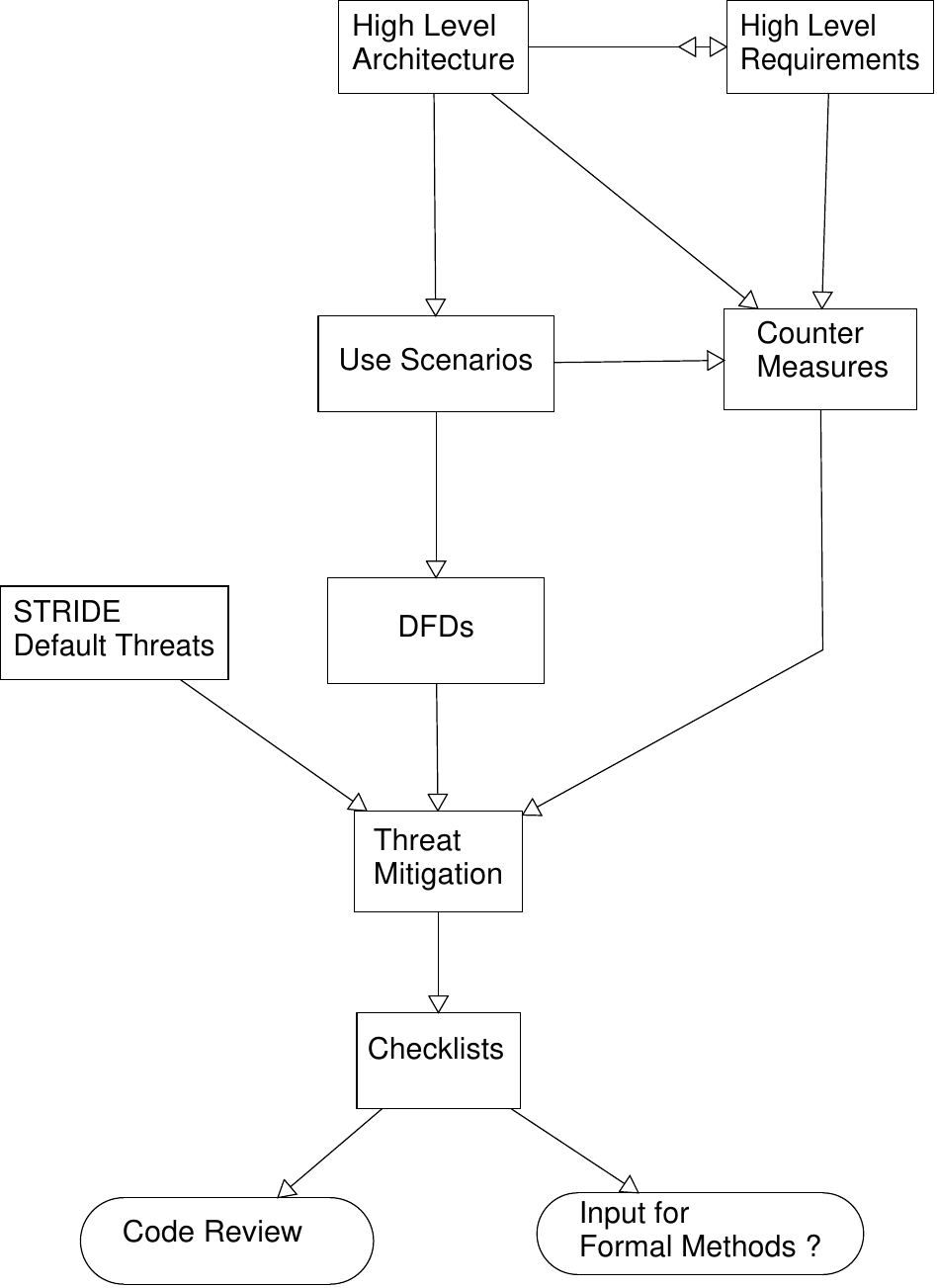}
    \caption{Workflow for modified STRIDE Analysis}
    \label{fig:Workflow for modified STRIDE Analysis}
\end{figure}

Figure \ref{fig:Workflow for modified STRIDE Analysis} shows the modified workflow
for the performed STRIDE analysis. Use Scenarios were developed on base of the
high level architecture described in  \cite{kuntze:fuchs:rudolph:2010}.

\begin{table}[H]
\caption{Mapping STRIDE Threats to DFD Element Types}
\label{tab:Mapping STRIDE Threats to DFD Element Types}
\begin{tabular}{lcccccc}

\hline
 {\bf DFD Element Type}  & \hspace{0.5cm} {\bf S}  & \hspace{0.5cm} {\bf T} & \hspace{0.5cm} {\bf R}   & \hspace{0.5cm} {\bf I}
 & \hspace{0.5cm} {\bf D } & \hspace{0.5cm} {\bf E } \\
\hline
External Entity & \hspace{0.5cm} X & \hspace{0.5cm}   & \hspace{0.5cm} X &
\hspace{0.5cm}   & \hspace{0.5cm}   & \hspace{0.5cm}   \\
  \\
Data Flow       & \hspace{0.5cm}   & \hspace{0.5cm} X & \hspace{0.5cm}   & \hspace{0.5cm} X & \hspace{0.5cm} X & \hspace{0.5cm}   \\
Data Store      & \hspace{0.5cm}   & \hspace{0.5cm} X & \hspace{0.5cm} X & \hspace{0.5cm} X & \hspace{0.5cm} X & \hspace{0.5cm}   \\
Process         & \hspace{0.5cm} X & \hspace{0.5cm} X & \hspace{0.5cm} X & \hspace{0.5cm} X & \hspace{0.5cm} X & \hspace{0.5cm} X \\
\hline
\end{tabular} 
\end{table}

\subsubsection{External Dependencies}\label{sec:External Dependencies}
Components of the system running on NaDa nodes
\begin{itemize}
  \item A TPM crypto processor will be used on NaDa nodes.
  \item Ubuntu hypervisor running Xen will be used as virtualization technology
    to run NaDa management and App Slices.
  \item The sHype \cite{shype:2005} hypervisor security  architecture will be used to control
    information flow between App Slices sharing a single NaDa Node.
  \item Authentication and Attestation, thus the realization  of the overlay
    network, and the NaDa P2P protocol will be based on an implementation of a
    trusted P2P protocol described in \cite{kuntze:fuchs:rudolph:2010}.
  \item Beside his own implementation of a P2P protocol the customer has to
    implement an interface to be able to receive commands from Node Management,
    and to deliver \ndt{APP\_Content}:
     \begin{itemize}
        \item \ndt{APP\_User\_Request} for user interaction
        \item \ndt{APP\_Log} to receive accounting information from Node Management
     \end{itemize}
\end{itemize}

% \newpage
\subsubsection{DFDs}\label{sec:DFDs}

\begin{figure}[H]
    \centering
    \includegraphics[scale=0.7]{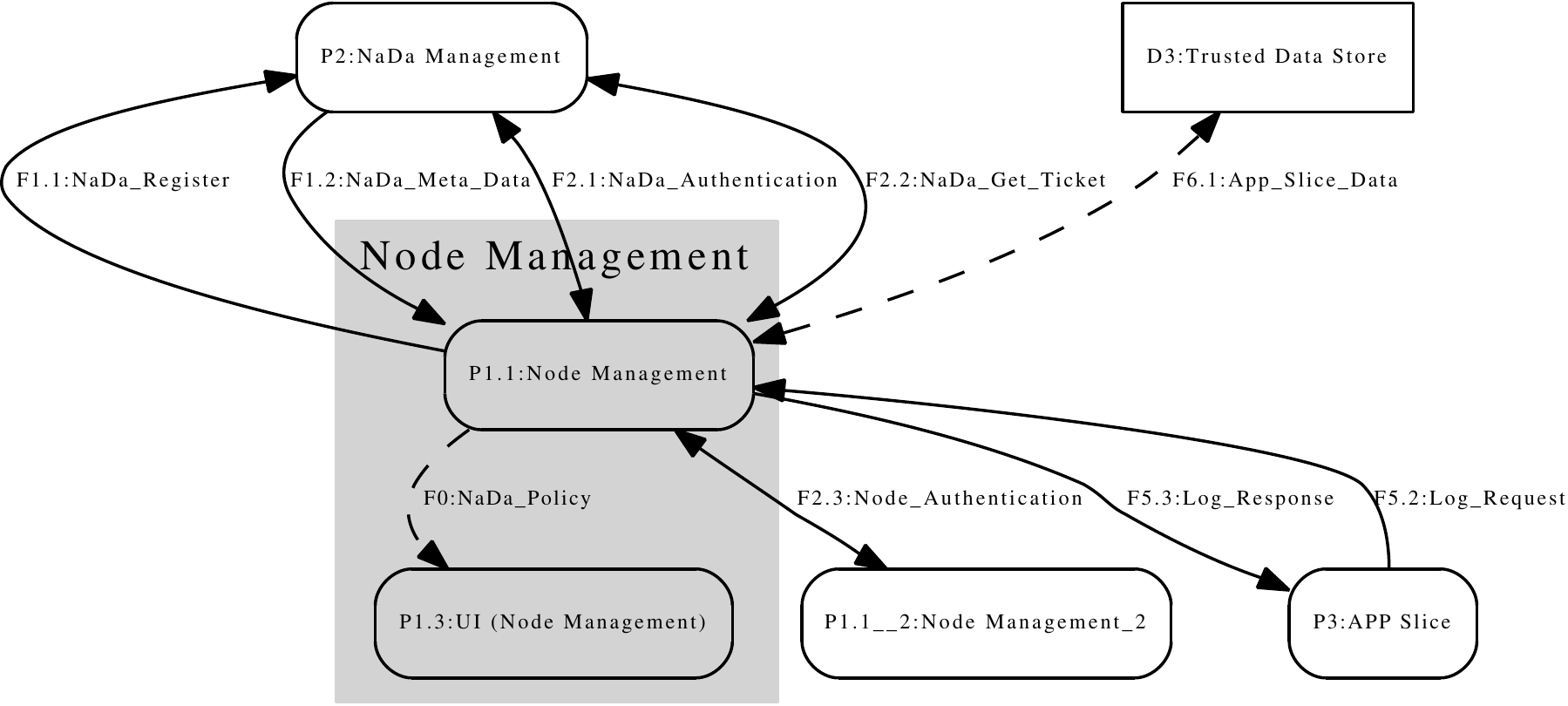}
    \includegraphics[scale=0.7]{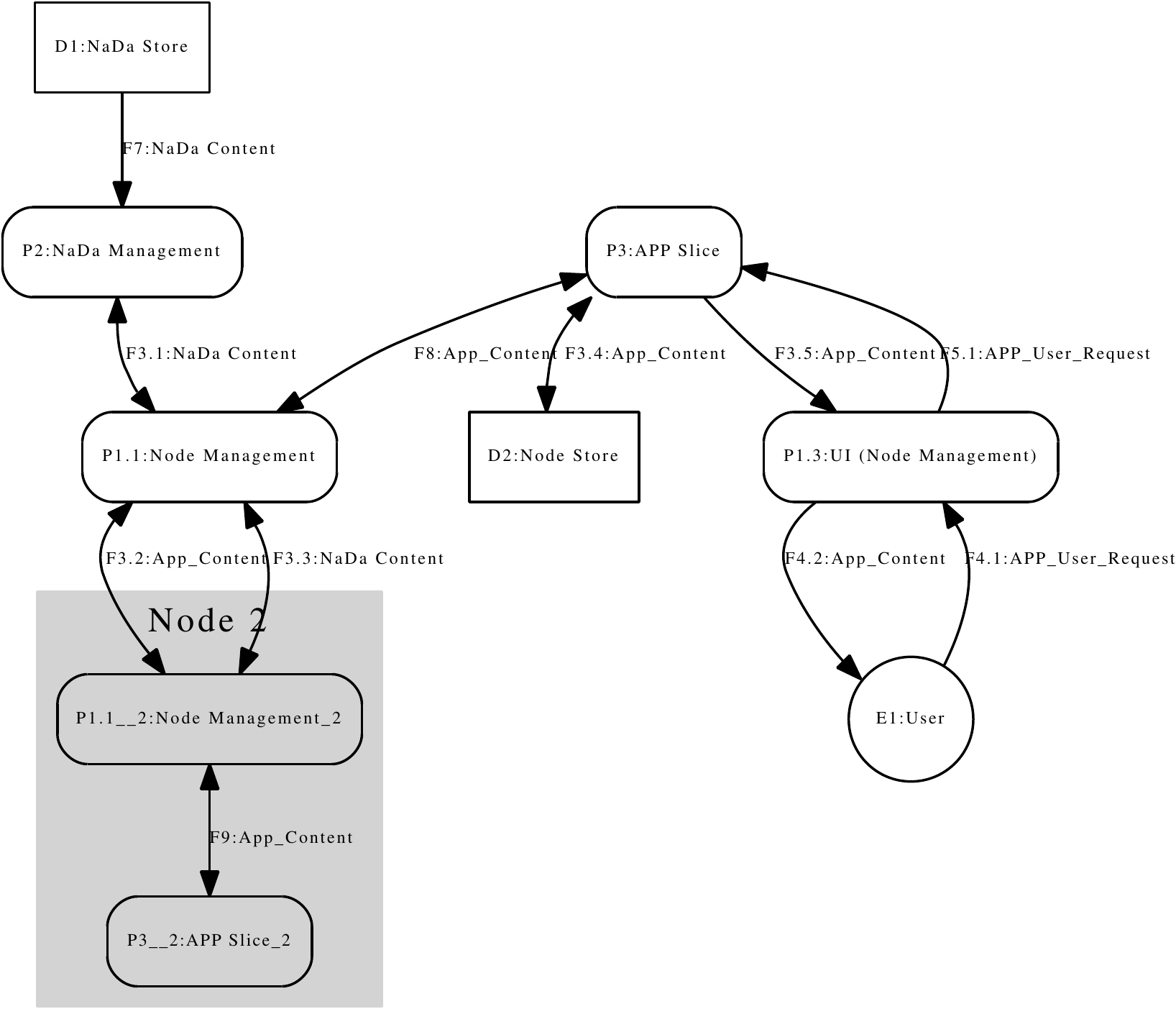}
    \caption{DFDs}
    \label{fig:DFD}
\end{figure}

All primitives and data exchanged between NaDa components are included in the 
following data flow diagrams (DFD). Data flows not going through trust boundaries
are  depicted with dashed arrows. For all other data flows trust
boundaries are crossed. Trust boundaries exist for data flows between
components with different privileges. E.g. an App Slice has lower privileges
than Node Management. To achieve better clarity in Figure \ref{fig:DFD} the DFD is
divided into two subdiagramms.

%\subsection{Primitives}

\setlongtables \begin{longtable}{llX}
\caption{DFD Elements} \label{tab:DFD Elements} \\
\hline
External Entities   & E1  & User \\

\hline
Processes   & P1.1  & Node Management \\
    & P1.2  & Node Monitoring \\
    & P1.3  & UI (Node Management) \\
    & P2  & NaDa Management \\
    & P3  & APP Slice \\

\hline
Data Stores   & D1  & NaDa Store \\
    & D2  & Node Store \\
    & D3  & Trusted Data Store \\

\hline
Data Flows   & F0  & NaDa\_Policy  P1.1 $\rightarrow$ P1.3 \\
    & F1.1  & NaDa\_Register  P1.1 $\rightarrow$ P2 \\
    & F1.2  & NaDa\_Meta\_Data  P2 $\rightarrow$ P1.1 \\
    & F2.1  & NaDa\_Authentication  P2 $\leftrightarrow$ P1.1 \\
    & F2.2  & NaDa\_Get\_Ticket  P2 $\leftrightarrow$ P1.1 \\
    & F2.3  & Node\_Authentication  P1.1 $\leftrightarrow$ P1.1\_\_2 \\
    & F3.1  & NaDa Content  P2 $\leftrightarrow$ P1.1 \\
    & F3.2  & App\_Content  P1.1 $\leftrightarrow$ P1.1\_\_2 \\
    & F3.3  & NaDa Content  P1.1 $\leftrightarrow$ P1.1\_\_2 \\
    & F3.4  & App\_Content  P3 $\leftrightarrow$ D2 \\
    & F3.5  & App\_Content  P3 $\rightarrow$ P1.3 \\
    & F4.1  & APP\_User\_Request  E1 $\rightarrow$ P1.3 \\
    & F4.2  & App\_Content  P1.3 $\rightarrow$ E1 \\
    & F5.1  & APP\_User\_Request  P1.3 $\rightarrow$ P3 \\
    & F5.2  & Log\_Request  P3 $\rightarrow$ P1.1 \\
    & F5.3  & Log\_Response  P1.1 $\rightarrow$ P3 \\
    & F6.1  & App\_Slice\_Data  D3 $\leftrightarrow$ P1.1 \\
    & F7  & NaDa Content  D1 $\rightarrow$ P2 \\
    & F8  & App\_Content  P3 $\leftrightarrow$ P1.1 \\
    & F9  & App\_Content  P1.1\_\_2 $\leftrightarrow$ P3\_\_2 \\

\end{longtable}

\setlongtables \begin{longtable}{lX}
\caption{Threats to the System} \label{tab:Threats to the System} \\
\hline{\bf Threat Type} & {\bf DFD Item Number} \endhead
\hline
Spoofing  & External Entities: E1 \\
 & Processes: P1.1, P1.2, P1.3, P2, P3 \\
 & Data Flows: F0, F1.1, F1.2, F2.1, F2.2, F2.3, F3.1, F3.2, F3.3, F3.4, F3.5, F4.1, F4.2, F5.1, F5.2, F5.3, F7, F8 \\

\hline
Tampering  & Processes: P1.1, P1.2, P1.3, P2, P3 \\
 & Data Stores: D1, D2, D3 \\
 & Data Flows: F0, F1.1, F1.2, F2.1, F2.2, F2.3, F3.1, F3.2, F3.3, F3.4, F3.5, F4.1, F4.2, F5.1, F5.2, F5.3, F6.1, F7, F8, F9 \\

\hline
Repudiation  & Data Flows: F3.4, F3.5, F4.1, F4.2, F5.1, F5.2, F5.3, F8 \\

\hline
Information Disclosure  & Processes: P1.1, P1.2, P1.3, P2, P3 \\
 & Data Stores: D1, D2, D3 \\
 & Data Flows: F0, F1.1, F1.2, F2.1, F2.2, F2.3, F3.1, F3.2, F3.3, F3.4, F3.5, F4.1, F4.2, F5.1, F5.2, F5.3, F6.1, F7, F8, F9 \\

\hline
EoP  & Processes: P3 \\

\hline
\end{longtable} 

\paragraph{Plan Mitigations}
Table \ref{tab:High Level Mitigation Techniques} lists high level mitigation
techniques proposed in \cite{kuntze:fuchs:rudolph:2010}:

\begin{table}[H]
\caption{High Level Mitigation Techniques}
\label{tab:High Level Mitigation Techniques}
\begin{tabularx}{\textwidth}{ll}

\hline
 {\bf Threat Typ}  & {\bf Mitigation Technique} \hspace{\textwidth}\\
\hline
 Spoofing & Authentication \\
 Tampering & Integrity \\
 Repudiation & Non-repudiation services \\
 Information disclosure & Confidentiality \\
 DoS & Availability \\
 EoP & Authorization \\
\hline
\end{tabularx} 
\end{table}
In \cite{kuntze:repp:2009} mitigation techniques provided by the NaDa platform were
presented in textual form.  In the next step the measures described in this
paper were divided into ``atomic''  measures. This partitioning allows
to assign single  measures to STRIDE threats and in the next step to the
table where DFD elements are assigned to these steps. 
Isolation of App slices and the access control to physical resources will be
assured by virtualization. Virtualization is not listed as specific mitigation
technique in the following table. Only measures necessary to control interfaces
of the virtual images and to enforce the NaDa policies will be listed.
Also Trusted Platform Module will not be listed as an ``atomic'' mitigation
technique.

\setlongtables \begin{longtable}{lX}
\caption{NaDa Measures} \label{tab:NaDa Measures} \\
\hline
\multicolumn{2} {l} {\bf Measures} \endhead
\hline
M0.1 & ISP responsibility \\
M0.2 & Customer responsibility \\
M0.3 & ** No measure ** \\
M1 & Mandatory access control in hypervisor (sHype) \\
M2 & Firewall rules running in the NaDa nodes \\
M3 & Trusted ticket for access to overlay net \\
M4.0 & Create signed log entry with trusted timestamp \\
M4.1 & Send log entry for User Action to APP slice \\
M4.2 & Write log entry for Action to Trusted Data Store \\
M4.3 & Send log entry for Action to NaDa Management \\
M5 & Check fingerprint of APP Slice \\
M6.1 & Trusted Boot, initialize measuring data for loaded code. \\
M6.2 & Initialize Trusted Data Store \\
M6.3 & Initialize Node Store \\
M7.1 & Encrypt Trusted Data Store \\
M7.2 & Encrypt Node Stores \\
M8 & ISP Certificate \\
M9 & Remote node attestation \\
M10 & Mutual authentication \\
M11 & Meta data signed by ISP \\
M12 & Encryption of channel \\
M13 & Encryption Overlay Net \\
M14 & App Identifier displayed by ISP \\
M15 & Data stored in Trusted Data Store \\
M16 & e.g. HDMI digital content protection \\
M17 & e.g. Digital watermarking \\
M18 & Don't store data worthy to be protected in this entity \\
M19 & Correct assignment of Resource\_ID to AppSlice \\
\hline
\end{longtable} 

\begin{description}

\item[M1] sHype stands for secure hypervisor. sHype is a hypervisor security
architecture developed by IBM Research available for XEN virtualization.
sHype supports policy-based information flow control for virtual machines. Two
standard policies "Chinese Wall" and "Simple Type Enforcement" are
available. It's possible to define coalitions which can share resources. sHype
types are used in these policies and define the granularity for 
App Slice coalitions and the available resources (App Stores).  NaDa policies must define
the mapping of NaDa\_Resource\_IDs to these types. For mandatory access
control to App Stores simple type enforcement policy will be used (App Slices
can use resources with equal types). So according to customer policies isolation
from resources of other customers can be enforced, and the sharing of resources 
between different App Slices can be permitted. The policy defined by the
customer has to be verified by the ISP. Only certified policies will be used to
configure access rights of App slices.

\item[M2] Virtualization is used to isolate App Slices according to policies
defined by the customers. The network connections of APP Slices are critical
parts of the architecture respective to breaking these policies. A standard framework
(e.g. iptables) has to be used to control network traffic between different App
Slices by the privileged domain of the NaDa Node. The default firewall rules
have to enforce strict separation of App Slices. Customer policies allowing
communication of different App Slices on the same node have to be certified by the ISP.

\item[M3]  There are two types of "overlay nets" . The overlay net which is used to
exchange $Nada\_Content$ ( App Slices, updates) and the overlay nets for customer
applications. Beside the IP addresses the NaDa resource identifier (see
\ref{sec:Used Terms}) is used to address entities in the NaDa Network and to define the
structure of the overlay net.  Only one specific resource identifier is used to
address NaDa entities in the network. Thus there is one NaDa specific overlay
net which can be used by all NaDa entities for maintenance purposes.  The NaDa
resource identifier must be different from customer resource identifiers. The
customer defines the policy which other applications (identified by their
\ndt{NaDa\_Resource\_ID}) can use a certain resource (App\_Slice). This policy is
delivered to NaDa nodes together with App Slice images. 
Thus this policy implicitly defines the structure of the overlay net and also
has to be verified and certified by the ISP before installation.
Base for the communication in the overlay net is the ``trusted ticket'' computed
in the last step of the trusted P2P protocol presented in
\cite{kuntze:fuchs:rudolph:2010}. Additional to the verification of the ticket 
the policy for accessing NaDa resources has to be checked with the
NaDa\_Resource\_ID which is part of the ticket.
 A symetric key for encryption can be computed after the ticket
is delivered to the addressed node and the remote attestation of this node is
performed successfully.

\item[M4.0] A non-repudiation service must exist to produce trusted
  log data with timestamps. Trusted log would not possible without performing
  time  synchronization during boot process. Time synchronization must be
  repeated in certain time intervals to avoid inaccuracies. Log entries have
  to be signed with a key derived from the root of trust of the NaDa Node.

\item[M4.1] Log entries collected by the ISP have to be created according to
  M4.0. Depending on log policy. The log entries  must be written to Trusted Data
  Store or have to be sent to NaDa Monitoring. 

\item[M4.2, M4.3] Log entries collected by the ISP have to be created according to
  M4.0. Depending on log policy The log entries  must be written to Trusted Data
  Store or have to be sent to NaDa Monitoring. 

\item[M5] The fingerprints of App Slices have to be checked before
  installation. The fingerprint is part of the meta file signed by ISP which
  describes the NaDa Content to be downloaded.

\item[M6.1] Every code running on a NaDa Node has to be measured before execution. (for
measurement of App Slice code see M5).  NaDa Nodes have to be equipped with
platform firmware that must be implicitly trusted. This firmware has to initialize the
root of trust for reporting (RTR), which accumulates measuring data for loaded
code. For other entities of the NaDa architecture it must be possible to check
the attestation for this measuring data to verify the integrity of the loaded
code.

\item[M6.2] The key for encryption of the Trusted Data Store has to be
  computed on the node and must be bound to the attested state of the NaDa Node. Thus there
  must exist a root of trust for storage and the possibility to bind usage of
  stored data do measuring data computed in M6.1. Usage of this key is only
  permitted in an attested state.

\item[M6.3] A key for encryption of the Node Store has to be computed.
  This key has to be stored in Trusted  Data Store under the Resource\_ID 
  of the corresponding App Slice. So this key 
  is implicitly bound to an attested state of the platform. Access to Node Store
  is only possible with attested software.

\item[M7.1] Trusted data store has to be encrypted to guarantee integrity of
  stored policy data, NaDa Management data, and log data.

\item[M7.2] Data of Node Stores assigned to certain App Slices has to be
  encrypted to prevent illegal access to content and to prevent manipulation of
  customer content.

\item[M8] The ISP certificate is part of the NaDa Management software installed on the
  NaDa node before delivery in a trustworthy environment.

\item[M9] Attestation of the actual node configuration and the actual firmware
  version to remote parties must be possible. Measuring data computed in M6.1
  will be used for remote attestation. 

\item[M10] Mutual authentication between Node Management and NaDa Management
  must be provided. In the case of centralized NaDa Management the protocol
  described in  MSC Figure 2  of the trusted P2P protocol is used. Parts of the
  NaDa Management could run distributed. In this case the protocol from MSC
  Figure 3 must be used. Base for the mutual authentication are the remote
  attestation capabilities of the nodes (M9) and the ISP certificate used to 
  verify messages from the centralized NaDa Management (M8). A symmetric key for 
  encryption will be exchanged when this protocol is used.

\item[M11] 
  The metafile describing NaDa specific content has to be signed by ISP tu
  guarantee integrity of this content.

\item[M12] 
  Network traffic between centralized NaDa Management and NaDa Nodes has to be
  encrypted. The symmetric key computed in M9 is used for encryption.

\item[M13]
  Network traffic between NaDa Nodes has to be
  encrypted. The symmetric key computed in M3 is used for encryption.

\item[M14] To avoid spoofing of the user interface for certain App Slices the
  ISP should ensure that the identity of the current content provider is
  presented for the user, beside the presentation of the data provided by the
  content provider.

\item[M15] All data used by NaDa Management and NaDa Monitoring has to be stored in 
  Trusted Data Store, to ensure integrity of this data. So access to data is
  possible for previously defined software only (M6.1, M6.2).

\item[M16] For content delivery digital content protection (e.g. HDMI) could be
  used.

\item[M17] NaDa will provide capability for digital watermarking in the
  standard libraries of App Slices for the content providers.

\item[M18] If a entity is not encrypted do not store secret data in this entity.

\item[M19] Correct assignment of \ndt{NaDa\_Resource\_IDs} to App Slices and to
  physical resources is essential for correct isolation of App Slices. Parts of
  the software performing this assignment should be reviewed, or if possible
  investigated using formal methods.

\end{description}

\begin{figure}[H]
    \centering
    \includegraphics[scale=0.6]{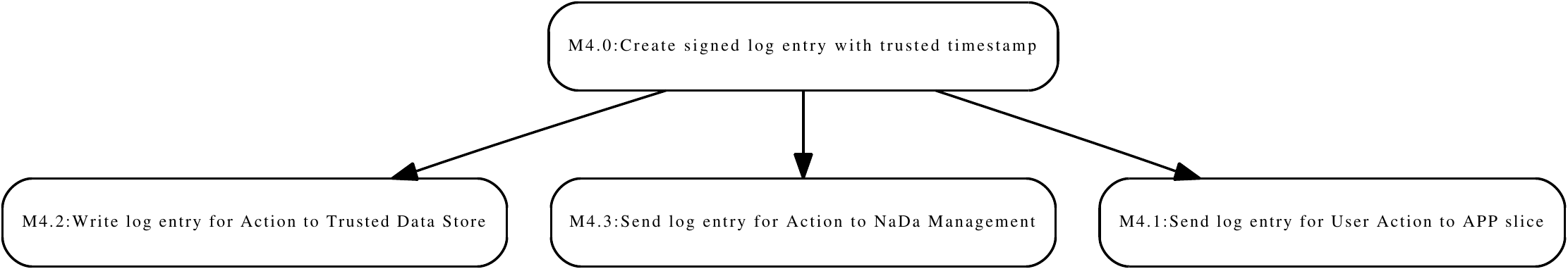}
    \vspace{0.2cm}
    \includegraphics[scale=0.6]{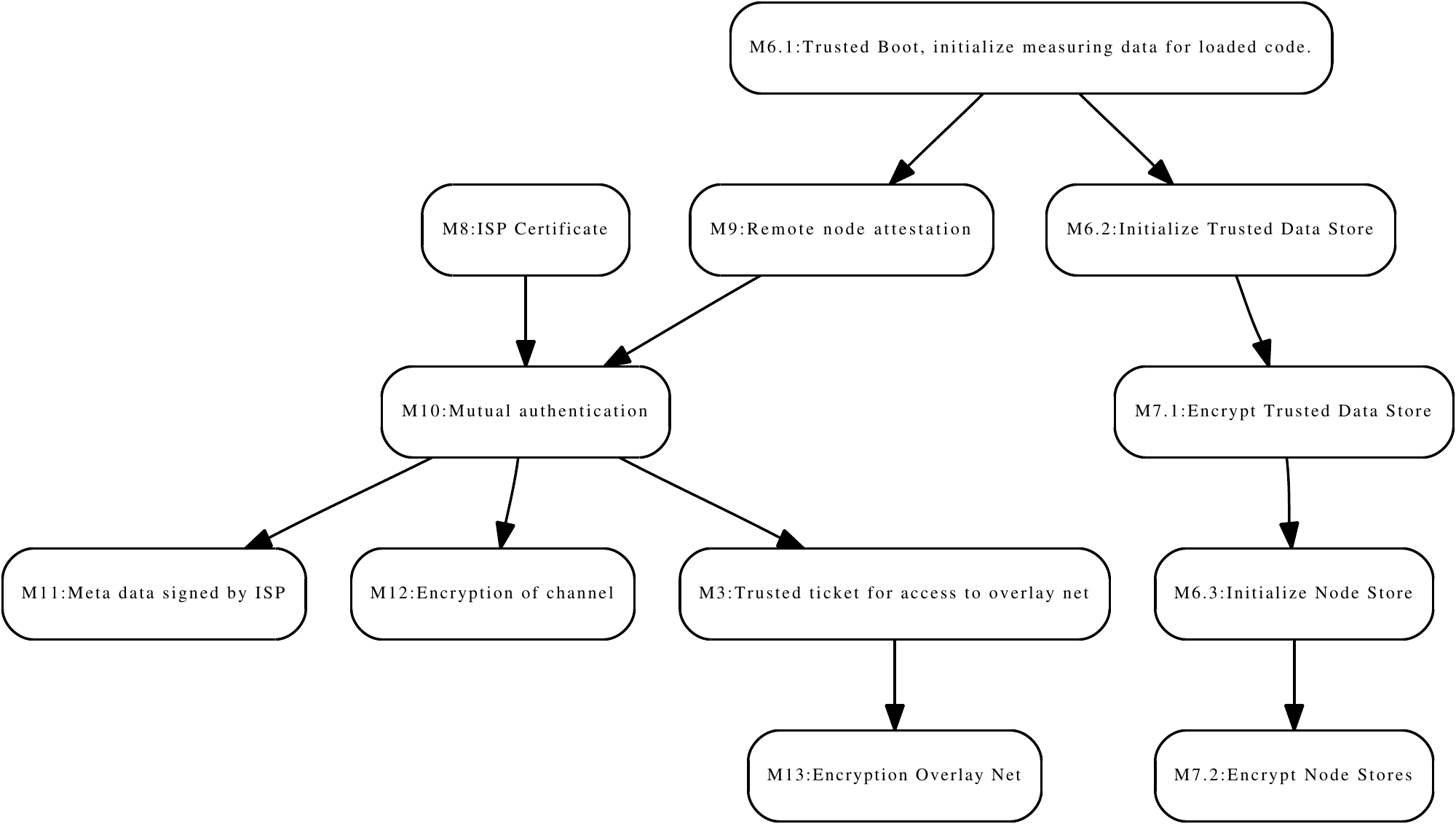}
    \caption{Measure Dependency}
    \label{fig:Measure Dependency}
\end{figure}

All tables presented in this paper are generated from YAML files
(http//www.yaml.org). 
YAML is a human friendly data serialization  standard for many
programming languages. This offers several opportunities:
\begin{itemize}
 \item Data could be used to generate specifications for model checking.
 \item Data could be used for generation of check-lists.
 \item Data could be used for policy generation.
\end{itemize}
Also dependencies between single measures are defined in the YAML files (see Figure 
\ref{fig:Measure Dependency}).

Table \ref{tab:NaDa Mitigation Techniques} assigns mitigation techniques
provided by the NaDa platform to the high level mitigation
table \ref{tab:High Level Mitigation Techniques}. 
techniques proposed in \cite{Howard:Lipner:2006}.

\setlongtables \begin{longtable}{lX}
\caption{NaDa Mitigation Techniques} \label{tab:NaDa Mitigation Techniques} \\
\hline
\multicolumn{2} {l} {{\bf Mitigation Technique} and Measures} \endhead
\hline
\multicolumn{2} {l} {{\bf Authentication (Spoofing) }}\\
M0.2 & Customer responsibility \\
M8 & ISP Certificate \\
M2 & Firewall rules running in the NaDa nodes \\
M3 & Trusted ticket for access to overlay net \\
M9 & Remote node attestation \\
M14 & App Identifier displayed by ISP \\
M10 & Mutual authentication \\
&- F2.1 provides M10 \\
M12 & Encryption of channel \\
M13 & Encryption Overlay Net \\
M1 & Mandatory access control in hypervisor (sHype) \\
M19 & Correct assignment of Resource\_ID to AppSlice \\
M0.3 & ** No measure ** \\
M0.1 & ISP responsibility \\
\hline
\multicolumn{2} {l} {{\bf Integrity (Tampering) }}\\
M0.1 & ISP responsibility \\
M7.2 & Encrypt Node Stores \\
M7.1 & Encrypt Trusted Data Store \\
M6.1 & Trusted Boot, initialize measuring data for loaded code. \\
M5 & Check fingerprint of APP Slice \\
M9 & Remote node attestation \\
M10 & Mutual authentication \\
&- F2.1 provides M10 \\
M12 & Encryption of channel \\
M11 & Meta data signed by ISP \\
&- F2.3 provides M11 \\
M13 & Encryption Overlay Net \\
M3 & Trusted ticket for access to overlay net \\
M1 & Mandatory access control in hypervisor (sHype) \\
M19 & Correct assignment of Resource\_ID to AppSlice \\
M0.3 & ** No measure ** \\
M4.0 & Create signed log entry with trusted timestamp \\
M2 & Firewall rules running in the NaDa nodes \\
\hline
\multicolumn{2} {l} {{\bf Non repudiaton service (Repudiation) }}\\
M17 & e.g. Digital watermarking \\
M4.1 & Send log entry for User Action to APP slice \\
M4.0 & Create signed log entry with trusted timestamp \\
M4.3 & Send log entry for Action to NaDa Management \\
\hline
\multicolumn{2} {l} {{\bf Confidentiality (Information Disclosure) }}\\
M7.2 & Encrypt Node Stores \\
M7.1 & Encrypt Trusted Data Store \\
M18 & Don't store data worthy to be protected in this entity \\
M0.1 & ISP responsibility \\
M0.2 & Customer responsibility \\
M12 & Encryption of channel \\
M10 & Mutual authentication \\
&- F2.1 provides M10 \\
M13 & Encryption Overlay Net \\
M2 & Firewall rules running in the NaDa nodes \\
M19 & Correct assignment of Resource\_ID to AppSlice \\
M16 & e.g. HDMI digital content protection \\
\hline
\multicolumn{2} {l} {{\bf Authorization (EoP) }}\\
M1 & Mandatory access control in hypervisor (sHype) \\
M2 & Firewall rules running in the NaDa nodes \\
M3 & Trusted ticket for access to overlay net \\
\hline
\end{longtable}

\setlongtables \begin{longtable}{lX}
\caption{Assigned Countermeasures} \label{tab:Assigned Countermeasures} \\
\hline{\bf Threat Type} & {\bf DFD Item Number} \endhead
\hline
\begin{minipage}{3cm} Spoofing \end{minipage}
 & {\bf External Entities:} \\
 & E1: User \\
 & - M0.2 : Customer responsibility \\
 & {\bf Processes:} \\
 & P1.1: Node Management \\
 & - M9 : Remote node attestation \\
 & \hspace*{0.25cm} - M6.1 : Trusted Boot, initialize measuring data for loaded code. \\
 & - M3 : Trusted ticket for access to overlay net \\
 & \hspace*{0.25cm} - M10 : Mutual authentication \\
 & \hspace*{0.25cm} \hspace*{0.25cm} - M8 : ISP Certificate \\
 & \hspace*{0.25cm} \hspace*{0.25cm} - M9 : Remote node attestation \\
 & \hspace*{0.25cm} \hspace*{0.25cm} \hspace*{0.25cm} - M6.1 : Trusted Boot, initialize measuring data for loaded code. \\
 & P1.2: Node Monitoring \\
 & P1.3: UI (Node Management) \\
 & - M14 : App Identifier displayed by ISP \\
 & P2: NaDa Management \\
 & - M8 : ISP Certificate \\
 & P3: APP Slice \\
 & - M2 : Firewall rules running in the NaDa nodes \\
 & - M3 : Trusted ticket for access to overlay net \\
 & \hspace*{0.25cm} - M10 : Mutual authentication \\
 & \hspace*{0.25cm} \hspace*{0.25cm} - M8 : ISP Certificate \\
 & \hspace*{0.25cm} \hspace*{0.25cm} - M9 : Remote node attestation \\
 & \hspace*{0.25cm} \hspace*{0.25cm} \hspace*{0.25cm} - M6.1 : Trusted Boot, initialize measuring data for loaded code. \\
 & {\bf Data Flows:} \\
 & F1.1 - NaDa\_Register: Node Management $\rightarrow$ NaDa Management \\
 & - M10 : Mutual authentication \\
 & \hspace*{0.25cm} - M8 : ISP Certificate \\
 & \hspace*{0.25cm} - M9 : Remote node attestation \\
 & \hspace*{0.25cm} \hspace*{0.25cm} - M6.1 : Trusted Boot, initialize measuring data for loaded code. \\
 & F1.2 - NaDa\_Meta\_Data: NaDa Management $\rightarrow$ Node Management \\
 & - M10 : Mutual authentication \\
 & \hspace*{0.25cm} - M8 : ISP Certificate \\
 & \hspace*{0.25cm} - M9 : Remote node attestation \\
 & \hspace*{0.25cm} \hspace*{0.25cm} - M6.1 : Trusted Boot, initialize measuring data for loaded code. \\
 & F2.1 - NaDa\_Authentication: NaDa Management $\leftrightarrow$ Node Management \\
 & - M8 : ISP Certificate \\
 & - M9 : Remote node attestation \\
 & \hspace*{0.25cm} - M6.1 : Trusted Boot, initialize measuring data for loaded code. \\
 & F2.2 - NaDa\_Get\_Ticket: NaDa Management $\leftrightarrow$ Node Management \\
 & - M10 : Mutual authentication \\
 & \hspace*{0.25cm} - M8 : ISP Certificate \\
 & \hspace*{0.25cm} - M9 : Remote node attestation \\
 & \hspace*{0.25cm} \hspace*{0.25cm} - M6.1 : Trusted Boot, initialize measuring data for loaded code. \\
 & F2.3 - Node\_Authentication: Node Management $\leftrightarrow$ Node Management\_2 \\
 & - M3 : Trusted ticket for access to overlay net \\
 & \hspace*{0.25cm} - M10 : Mutual authentication \\
 & \hspace*{0.25cm} \hspace*{0.25cm} - M8 : ISP Certificate \\
 & \hspace*{0.25cm} \hspace*{0.25cm} - M9 : Remote node attestation \\
 & \hspace*{0.25cm} \hspace*{0.25cm} \hspace*{0.25cm} - M6.1 : Trusted Boot, initialize measuring data for loaded code. \\
 & F3.1 - NaDa Content: NaDa Management $\leftrightarrow$ Node Management \\
 & - M12 : Encryption of channel \\
 & \hspace*{0.25cm} - M10 : Mutual authentication \\
 & \hspace*{0.25cm} \hspace*{0.25cm} - M8 : ISP Certificate \\
 & \hspace*{0.25cm} \hspace*{0.25cm} - M9 : Remote node attestation \\
 & \hspace*{0.25cm} \hspace*{0.25cm} \hspace*{0.25cm} - M6.1 : Trusted Boot, initialize measuring data for loaded code. \\
 & F3.2 - App\_Content: Node Management $\leftrightarrow$ Node Management\_2 \\
 & - M3 : Trusted ticket for access to overlay net \\
 & \hspace*{0.25cm} - M10 : Mutual authentication \\
 & \hspace*{0.25cm} \hspace*{0.25cm} - M8 : ISP Certificate \\
 & \hspace*{0.25cm} \hspace*{0.25cm} - M9 : Remote node attestation \\
 & \hspace*{0.25cm} \hspace*{0.25cm} \hspace*{0.25cm} - M6.1 : Trusted Boot, initialize measuring data for loaded code. \\
 & - M3 : Trusted ticket for access to overlay net \\
 & \hspace*{0.25cm} - M10 : Mutual authentication \\
 & \hspace*{0.25cm} \hspace*{0.25cm} - M8 : ISP Certificate \\
 & \hspace*{0.25cm} \hspace*{0.25cm} - M9 : Remote node attestation \\
 & \hspace*{0.25cm} \hspace*{0.25cm} \hspace*{0.25cm} - M6.1 : Trusted Boot, initialize measuring data for loaded code. \\
 & F3.3 - NaDa Content: Node Management $\leftrightarrow$ Node Management\_2 \\
 & - M13 : Encryption Overlay Net \\
 & \hspace*{0.25cm} - M3 : Trusted ticket for access to overlay net \\
 & \hspace*{0.25cm} \hspace*{0.25cm} - M10 : Mutual authentication \\
 & \hspace*{0.25cm} \hspace*{0.25cm} \hspace*{0.25cm} - M8 : ISP Certificate \\
 & \hspace*{0.25cm} \hspace*{0.25cm} \hspace*{0.25cm} - M9 : Remote node attestation \\
 & \hspace*{0.25cm} \hspace*{0.25cm} \hspace*{0.25cm} \hspace*{0.25cm} - M6.1 : Trusted Boot, initialize measuring data for loaded code. \\
 & F3.4 - App\_Content: APP Slice $\leftrightarrow$ Node Store \\
 & - M1 : Mandatory access control in hypervisor (sHype) \\
 & F3.5 - App\_Content: APP Slice $\rightarrow$ UI (Node Management) \\
 & - M19 : Correct assignment of Resource\_ID to AppSlice \\
 & F4.1 - APP\_User\_Request: User $\rightarrow$ UI (Node Management) \\
 & - M0.2 : Customer responsibility \\
 & F4.2 - App\_Content: UI (Node Management) $\rightarrow$ User \\
 & - M0.3 : ** No measure ** \\
 & F5.1 - APP\_User\_Request: UI (Node Management) $\rightarrow$ APP Slice \\
 & - M3 : Trusted ticket for access to overlay net \\
 & \hspace*{0.25cm} - M10 : Mutual authentication \\
 & \hspace*{0.25cm} \hspace*{0.25cm} - M8 : ISP Certificate \\
 & \hspace*{0.25cm} \hspace*{0.25cm} - M9 : Remote node attestation \\
 & \hspace*{0.25cm} \hspace*{0.25cm} \hspace*{0.25cm} - M6.1 : Trusted Boot, initialize measuring data for loaded code. \\
 & F5.2 - Log\_Request: APP Slice $\rightarrow$ Node Management \\
 & - M19 : Correct assignment of Resource\_ID to AppSlice \\
 & F5.3 - Log\_Response: Node Management $\rightarrow$ APP Slice \\
 & - M19 : Correct assignment of Resource\_ID to AppSlice \\
 & F7 - NaDa Content: NaDa Store $\rightarrow$ NaDa Management \\
 & - M0.1 : ISP responsibility \\
 & F8 - App\_Content: APP Slice $\leftrightarrow$ Node Management \\
 & - M2 : Firewall rules running in the NaDa nodes \\
\hline
\begin{minipage}{3cm} Tampering \end{minipage}
 & {\bf Processes:} \\
 & P1.1: Node Management \\
 & - M6.1 : Trusted Boot, initialize measuring data for loaded code. \\
 & P1.2: Node Monitoring \\
 & - M6.1 : Trusted Boot, initialize measuring data for loaded code. \\
 & P1.3: UI (Node Management) \\
 & - M6.1 : Trusted Boot, initialize measuring data for loaded code. \\
 & P2: NaDa Management \\
 & - M0.1 : ISP responsibility \\
 & P3: APP Slice \\
 & - M5 : Check fingerprint of APP Slice \\
 & - M9 : Remote node attestation \\
 & \hspace*{0.25cm} - M6.1 : Trusted Boot, initialize measuring data for loaded code. \\
 & {\bf Data Stores:} \\
 & D1: NaDa Store \\
 & - M0.1 : ISP responsibility \\
 & D2: Node Store \\
 & - M7.2 : Encrypt Node Stores \\
 & \hspace*{0.25cm} - M6.3 : Initialize Node Store \\
 & \hspace*{0.25cm} \hspace*{0.25cm} - M7.1 : Encrypt Trusted Data Store \\
 & \hspace*{0.25cm} \hspace*{0.25cm} \hspace*{0.25cm} - M6.2 : Initialize Trusted Data Store \\
 & \hspace*{0.25cm} \hspace*{0.25cm} \hspace*{0.25cm} \hspace*{0.25cm} - M6.1 : Trusted Boot, initialize measuring data for loaded code. \\
 & D3: Trusted Data Store \\
 & - M7.1 : Encrypt Trusted Data Store \\
 & \hspace*{0.25cm} - M6.2 : Initialize Trusted Data Store \\
 & \hspace*{0.25cm} \hspace*{0.25cm} - M6.1 : Trusted Boot, initialize measuring data for loaded code. \\
 & {\bf Data Flows:} \\
 & F1.1 - NaDa\_Register: Node Management $\rightarrow$ NaDa Management \\
 & - M10 : Mutual authentication \\
 & \hspace*{0.25cm} - M8 : ISP Certificate \\
 & \hspace*{0.25cm} - M9 : Remote node attestation \\
 & \hspace*{0.25cm} \hspace*{0.25cm} - M6.1 : Trusted Boot, initialize measuring data for loaded code. \\
 & - M12 : Encryption of channel \\
 & \hspace*{0.25cm} - M10 : Mutual authentication \\
 & \hspace*{0.25cm} \hspace*{0.25cm} - M8 : ISP Certificate \\
 & \hspace*{0.25cm} \hspace*{0.25cm} - M9 : Remote node attestation \\
 & \hspace*{0.25cm} \hspace*{0.25cm} \hspace*{0.25cm} - M6.1 : Trusted Boot, initialize measuring data for loaded code. \\
 & F1.2 - NaDa\_Meta\_Data: NaDa Management $\rightarrow$ Node Management \\
 & - M10 : Mutual authentication \\
 & \hspace*{0.25cm} - M8 : ISP Certificate \\
 & \hspace*{0.25cm} - M9 : Remote node attestation \\
 & \hspace*{0.25cm} \hspace*{0.25cm} - M6.1 : Trusted Boot, initialize measuring data for loaded code. \\
 & - M11 : Meta data signed by ISP \\
 & \hspace*{0.25cm} - M10 : Mutual authentication \\
 & \hspace*{0.25cm} \hspace*{0.25cm} - M8 : ISP Certificate \\
 & \hspace*{0.25cm} \hspace*{0.25cm} - M9 : Remote node attestation \\
 & \hspace*{0.25cm} \hspace*{0.25cm} \hspace*{0.25cm} - M6.1 : Trusted Boot, initialize measuring data for loaded code. \\
 & - M12 : Encryption of channel \\
 & \hspace*{0.25cm} - M10 : Mutual authentication \\
 & \hspace*{0.25cm} \hspace*{0.25cm} - M8 : ISP Certificate \\
 & \hspace*{0.25cm} \hspace*{0.25cm} - M9 : Remote node attestation \\
 & \hspace*{0.25cm} \hspace*{0.25cm} \hspace*{0.25cm} - M6.1 : Trusted Boot, initialize measuring data for loaded code. \\
 & F2.1 - NaDa\_Authentication: NaDa Management $\leftrightarrow$ Node Management \\
 & - M10 : Mutual authentication \\
 & \hspace*{0.25cm} - M8 : ISP Certificate \\
 & \hspace*{0.25cm} - M9 : Remote node attestation \\
 & \hspace*{0.25cm} \hspace*{0.25cm} - M6.1 : Trusted Boot, initialize measuring data for loaded code. \\
 & - M12 : Encryption of channel \\
 & \hspace*{0.25cm} - M10 : Mutual authentication \\
 & \hspace*{0.25cm} \hspace*{0.25cm} - M8 : ISP Certificate \\
 & \hspace*{0.25cm} \hspace*{0.25cm} - M9 : Remote node attestation \\
 & \hspace*{0.25cm} \hspace*{0.25cm} \hspace*{0.25cm} - M6.1 : Trusted Boot, initialize measuring data for loaded code. \\
 & F2.2 - NaDa\_Get\_Ticket: NaDa Management $\leftrightarrow$ Node Management \\
 & - M10 : Mutual authentication \\
 & \hspace*{0.25cm} - M8 : ISP Certificate \\
 & \hspace*{0.25cm} - M9 : Remote node attestation \\
 & \hspace*{0.25cm} \hspace*{0.25cm} - M6.1 : Trusted Boot, initialize measuring data for loaded code. \\
 & - M12 : Encryption of channel \\
 & \hspace*{0.25cm} - M10 : Mutual authentication \\
 & \hspace*{0.25cm} \hspace*{0.25cm} - M8 : ISP Certificate \\
 & \hspace*{0.25cm} \hspace*{0.25cm} - M9 : Remote node attestation \\
 & \hspace*{0.25cm} \hspace*{0.25cm} \hspace*{0.25cm} - M6.1 : Trusted Boot, initialize measuring data for loaded code. \\
 & F2.3 - Node\_Authentication: Node Management $\leftrightarrow$ Node Management\_2 \\
 & - M10 : Mutual authentication \\
 & \hspace*{0.25cm} - M8 : ISP Certificate \\
 & \hspace*{0.25cm} - M9 : Remote node attestation \\
 & \hspace*{0.25cm} \hspace*{0.25cm} - M6.1 : Trusted Boot, initialize measuring data for loaded code. \\
 & F3.1 - NaDa Content: NaDa Management $\leftrightarrow$ Node Management \\
 & - M10 : Mutual authentication \\
 & \hspace*{0.25cm} - M8 : ISP Certificate \\
 & \hspace*{0.25cm} - M9 : Remote node attestation \\
 & \hspace*{0.25cm} \hspace*{0.25cm} - M6.1 : Trusted Boot, initialize measuring data for loaded code. \\
 & - M11 : Meta data signed by ISP \\
 & \hspace*{0.25cm} - M10 : Mutual authentication \\
 & \hspace*{0.25cm} \hspace*{0.25cm} - M8 : ISP Certificate \\
 & \hspace*{0.25cm} \hspace*{0.25cm} - M9 : Remote node attestation \\
 & \hspace*{0.25cm} \hspace*{0.25cm} \hspace*{0.25cm} - M6.1 : Trusted Boot, initialize measuring data for loaded code. \\
 & - M12 : Encryption of channel \\
 & \hspace*{0.25cm} - M10 : Mutual authentication \\
 & \hspace*{0.25cm} \hspace*{0.25cm} - M8 : ISP Certificate \\
 & \hspace*{0.25cm} \hspace*{0.25cm} - M9 : Remote node attestation \\
 & \hspace*{0.25cm} \hspace*{0.25cm} \hspace*{0.25cm} - M6.1 : Trusted Boot, initialize measuring data for loaded code. \\
 & - M5 : Check fingerprint of APP Slice \\
 & F3.2 - App\_Content: Node Management $\leftrightarrow$ Node Management\_2 \\
 & - M13 : Encryption Overlay Net \\
 & \hspace*{0.25cm} - M3 : Trusted ticket for access to overlay net \\
 & \hspace*{0.25cm} \hspace*{0.25cm} - M10 : Mutual authentication \\
 & \hspace*{0.25cm} \hspace*{0.25cm} \hspace*{0.25cm} - M8 : ISP Certificate \\
 & \hspace*{0.25cm} \hspace*{0.25cm} \hspace*{0.25cm} - M9 : Remote node attestation \\
 & \hspace*{0.25cm} \hspace*{0.25cm} \hspace*{0.25cm} \hspace*{0.25cm} - M6.1 : Trusted Boot, initialize measuring data for loaded code. \\
 & F3.3 - NaDa Content: Node Management $\leftrightarrow$ Node Management\_2 \\
 & - M3 : Trusted ticket for access to overlay net \\
 & \hspace*{0.25cm} - M10 : Mutual authentication \\
 & \hspace*{0.25cm} \hspace*{0.25cm} - M8 : ISP Certificate \\
 & \hspace*{0.25cm} \hspace*{0.25cm} - M9 : Remote node attestation \\
 & \hspace*{0.25cm} \hspace*{0.25cm} \hspace*{0.25cm} - M6.1 : Trusted Boot, initialize measuring data for loaded code. \\
 & - M11 : Meta data signed by ISP \\
 & \hspace*{0.25cm} - M10 : Mutual authentication \\
 & \hspace*{0.25cm} \hspace*{0.25cm} - M8 : ISP Certificate \\
 & \hspace*{0.25cm} \hspace*{0.25cm} - M9 : Remote node attestation \\
 & \hspace*{0.25cm} \hspace*{0.25cm} \hspace*{0.25cm} - M6.1 : Trusted Boot, initialize measuring data for loaded code. \\
 & - M13 : Encryption Overlay Net \\
 & \hspace*{0.25cm} - M3 : Trusted ticket for access to overlay net \\
 & \hspace*{0.25cm} \hspace*{0.25cm} - M10 : Mutual authentication \\
 & \hspace*{0.25cm} \hspace*{0.25cm} \hspace*{0.25cm} - M8 : ISP Certificate \\
 & \hspace*{0.25cm} \hspace*{0.25cm} \hspace*{0.25cm} - M9 : Remote node attestation \\
 & \hspace*{0.25cm} \hspace*{0.25cm} \hspace*{0.25cm} \hspace*{0.25cm} - M6.1 : Trusted Boot, initialize measuring data for loaded code. \\
 & F3.4 - App\_Content: APP Slice $\leftrightarrow$ Node Store \\
 & - M1 : Mandatory access control in hypervisor (sHype) \\
 & F3.5 - App\_Content: APP Slice $\rightarrow$ UI (Node Management) \\
 & - M19 : Correct assignment of Resource\_ID to AppSlice \\
 & F4.1 - APP\_User\_Request: User $\rightarrow$ UI (Node Management) \\
 & - M0.3 : ** No measure ** \\
 & F4.2 - App\_Content: UI (Node Management) $\rightarrow$ User \\
 & - M0.3 : ** No measure ** \\
 & F5.1 - APP\_User\_Request: UI (Node Management) $\rightarrow$ APP Slice \\
 & - M0.3 : ** No measure ** \\
 & F5.2 - Log\_Request: APP Slice $\rightarrow$ Node Management \\
 & - M0.3 : ** No measure ** \\
 & F5.3 - Log\_Response: Node Management $\rightarrow$ APP Slice \\
 & - M4.0 : Create signed log entry with trusted timestamp \\
 & F7 - NaDa Content: NaDa Store $\rightarrow$ NaDa Management \\
 & - M12 : Encryption of channel \\
 & \hspace*{0.25cm} - M10 : Mutual authentication \\
 & \hspace*{0.25cm} \hspace*{0.25cm} - M8 : ISP Certificate \\
 & \hspace*{0.25cm} \hspace*{0.25cm} - M9 : Remote node attestation \\
 & \hspace*{0.25cm} \hspace*{0.25cm} \hspace*{0.25cm} - M6.1 : Trusted Boot, initialize measuring data for loaded code. \\
 & F8 - App\_Content: APP Slice $\leftrightarrow$ Node Management \\
 & - M2 : Firewall rules running in the NaDa nodes \\
 & F9 - App\_Content: Node Management\_2 $\leftrightarrow$ APP Slice\_2 \\
\hline
\begin{minipage}{3cm} Repudiation \end{minipage}
 & {\bf Data Flows:} \\
 & F3.4 - App\_Content: APP Slice $\leftrightarrow$ Node Store \\
 & F3.5 - App\_Content: APP Slice $\rightarrow$ UI (Node Management) \\
 & - M17 : e.g. Digital watermarking \\
 & F4.1 - APP\_User\_Request: User $\rightarrow$ UI (Node Management) \\
 & - M4.1 : Send log entry for User Action to APP slice \\
 & \hspace*{0.25cm} - M4.0 : Create signed log entry with trusted timestamp \\
 & F4.2 - App\_Content: UI (Node Management) $\rightarrow$ User \\
 & - M17 : e.g. Digital watermarking \\
 & F5.1 - APP\_User\_Request: UI (Node Management) $\rightarrow$ APP Slice \\
 & - M4.1 : Send log entry for User Action to APP slice \\
 & \hspace*{0.25cm} - M4.0 : Create signed log entry with trusted timestamp \\
 & F5.2 - Log\_Request: APP Slice $\rightarrow$ Node Management \\
 & - M4.0 : Create signed log entry with trusted timestamp \\
 & F5.3 - Log\_Response: Node Management $\rightarrow$ APP Slice \\
 & F8 - App\_Content: APP Slice $\leftrightarrow$ Node Management \\
 & - M4.3 : Send log entry for Action to NaDa Management \\
 & \hspace*{0.25cm} - M4.0 : Create signed log entry with trusted timestamp \\
\hline
\begin{minipage}{3cm} Information Disclosure \end{minipage}
 & {\bf Processes:} \\
 & P1.1: Node Management \\
 & - M18 : Don't store data worthy to be protected in this entity \\
 & - M18 : Don't store data worthy to be protected in this entity \\
 & P1.2: Node Monitoring \\
 & - M18 : Don't store data worthy to be protected in this entity \\
 & P1.3: UI (Node Management) \\
 & - M18 : Don't store data worthy to be protected in this entity \\
 & P2: NaDa Management \\
 & - M0.1 : ISP responsibility \\
 & P3: APP Slice \\
 & - M0.2 : Customer responsibility \\
 & {\bf Data Stores:} \\
 & D1: NaDa Store \\
 & - M7.2 : Encrypt Node Stores \\
 & \hspace*{0.25cm} - M6.3 : Initialize Node Store \\
 & \hspace*{0.25cm} \hspace*{0.25cm} - M7.1 : Encrypt Trusted Data Store \\
 & \hspace*{0.25cm} \hspace*{0.25cm} \hspace*{0.25cm} - M6.2 : Initialize Trusted Data Store \\
 & \hspace*{0.25cm} \hspace*{0.25cm} \hspace*{0.25cm} \hspace*{0.25cm} - M6.1 : Trusted Boot, initialize measuring data for loaded code. \\
 & D2: Node Store \\
 & - M7.2 : Encrypt Node Stores \\
 & \hspace*{0.25cm} - M6.3 : Initialize Node Store \\
 & \hspace*{0.25cm} \hspace*{0.25cm} - M7.1 : Encrypt Trusted Data Store \\
 & \hspace*{0.25cm} \hspace*{0.25cm} \hspace*{0.25cm} - M6.2 : Initialize Trusted Data Store \\
 & \hspace*{0.25cm} \hspace*{0.25cm} \hspace*{0.25cm} \hspace*{0.25cm} - M6.1 : Trusted Boot, initialize measuring data for loaded code. \\
 & D3: Trusted Data Store \\
 & - M7.1 : Encrypt Trusted Data Store \\
 & \hspace*{0.25cm} - M6.2 : Initialize Trusted Data Store \\
 & \hspace*{0.25cm} \hspace*{0.25cm} - M6.1 : Trusted Boot, initialize measuring data for loaded code. \\
 & {\bf Data Flows:} \\
 & F1.1 - NaDa\_Register: Node Management $\rightarrow$ NaDa Management \\
 & - M12 : Encryption of channel \\
 & \hspace*{0.25cm} - M10 : Mutual authentication \\
 & \hspace*{0.25cm} \hspace*{0.25cm} - M8 : ISP Certificate \\
 & \hspace*{0.25cm} \hspace*{0.25cm} - M9 : Remote node attestation \\
 & \hspace*{0.25cm} \hspace*{0.25cm} \hspace*{0.25cm} - M6.1 : Trusted Boot, initialize measuring data for loaded code. \\
 & F1.2 - NaDa\_Meta\_Data: NaDa Management $\rightarrow$ Node Management \\
 & - M12 : Encryption of channel \\
 & \hspace*{0.25cm} - M10 : Mutual authentication \\
 & \hspace*{0.25cm} \hspace*{0.25cm} - M8 : ISP Certificate \\
 & \hspace*{0.25cm} \hspace*{0.25cm} - M9 : Remote node attestation \\
 & \hspace*{0.25cm} \hspace*{0.25cm} \hspace*{0.25cm} - M6.1 : Trusted Boot, initialize measuring data for loaded code. \\
 & F2.1 - NaDa\_Authentication: NaDa Management $\leftrightarrow$ Node Management \\
 & - M12 : Encryption of channel \\
 & \hspace*{0.25cm} - M10 : Mutual authentication \\
 & \hspace*{0.25cm} \hspace*{0.25cm} - M8 : ISP Certificate \\
 & \hspace*{0.25cm} \hspace*{0.25cm} - M9 : Remote node attestation \\
 & \hspace*{0.25cm} \hspace*{0.25cm} \hspace*{0.25cm} - M6.1 : Trusted Boot, initialize measuring data for loaded code. \\
 & F2.2 - NaDa\_Get\_Ticket: NaDa Management $\leftrightarrow$ Node Management \\
 & - M12 : Encryption of channel \\
 & \hspace*{0.25cm} - M10 : Mutual authentication \\
 & \hspace*{0.25cm} \hspace*{0.25cm} - M8 : ISP Certificate \\
 & \hspace*{0.25cm} \hspace*{0.25cm} - M9 : Remote node attestation \\
 & \hspace*{0.25cm} \hspace*{0.25cm} \hspace*{0.25cm} - M6.1 : Trusted Boot, initialize measuring data for loaded code. \\
 & F2.3 - Node\_Authentication: Node Management $\leftrightarrow$ Node Management\_2 \\
 & - M10 : Mutual authentication \\
 & \hspace*{0.25cm} - M8 : ISP Certificate \\
 & \hspace*{0.25cm} - M9 : Remote node attestation \\
 & \hspace*{0.25cm} \hspace*{0.25cm} - M6.1 : Trusted Boot, initialize measuring data for loaded code. \\
 & F3.1 - NaDa Content: NaDa Management $\leftrightarrow$ Node Management \\
 & - M12 : Encryption of channel \\
 & \hspace*{0.25cm} - M10 : Mutual authentication \\
 & \hspace*{0.25cm} \hspace*{0.25cm} - M8 : ISP Certificate \\
 & \hspace*{0.25cm} \hspace*{0.25cm} - M9 : Remote node attestation \\
 & \hspace*{0.25cm} \hspace*{0.25cm} \hspace*{0.25cm} - M6.1 : Trusted Boot, initialize measuring data for loaded code. \\
 & F3.2 - App\_Content: Node Management $\leftrightarrow$ Node Management\_2 \\
 & - M13 : Encryption Overlay Net \\
 & \hspace*{0.25cm} - M3 : Trusted ticket for access to overlay net \\
 & \hspace*{0.25cm} \hspace*{0.25cm} - M10 : Mutual authentication \\
 & \hspace*{0.25cm} \hspace*{0.25cm} \hspace*{0.25cm} - M8 : ISP Certificate \\
 & \hspace*{0.25cm} \hspace*{0.25cm} \hspace*{0.25cm} - M9 : Remote node attestation \\
 & \hspace*{0.25cm} \hspace*{0.25cm} \hspace*{0.25cm} \hspace*{0.25cm} - M6.1 : Trusted Boot, initialize measuring data for loaded code. \\
 & F3.3 - NaDa Content: Node Management $\leftrightarrow$ Node Management\_2 \\
 & - M13 : Encryption Overlay Net \\
 & \hspace*{0.25cm} - M3 : Trusted ticket for access to overlay net \\
 & \hspace*{0.25cm} \hspace*{0.25cm} - M10 : Mutual authentication \\
 & \hspace*{0.25cm} \hspace*{0.25cm} \hspace*{0.25cm} - M8 : ISP Certificate \\
 & \hspace*{0.25cm} \hspace*{0.25cm} \hspace*{0.25cm} - M9 : Remote node attestation \\
 & \hspace*{0.25cm} \hspace*{0.25cm} \hspace*{0.25cm} \hspace*{0.25cm} - M6.1 : Trusted Boot, initialize measuring data for loaded code. \\
 & F3.4 - App\_Content: APP Slice $\leftrightarrow$ Node Store \\
 & - M2 : Firewall rules running in the NaDa nodes \\
 & F3.5 - App\_Content: APP Slice $\rightarrow$ UI (Node Management) \\
 & - M19 : Correct assignment of Resource\_ID to AppSlice \\
 & F4.1 - APP\_User\_Request: User $\rightarrow$ UI (Node Management) \\
 & - M16 : e.g. HDMI digital content protection \\
 & F4.2 - App\_Content: UI (Node Management) $\rightarrow$ User \\
 & - M16 : e.g. HDMI digital content protection \\
 & F5.1 - APP\_User\_Request: UI (Node Management) $\rightarrow$ APP Slice \\
 & - M2 : Firewall rules running in the NaDa nodes \\
 & F5.2 - Log\_Request: APP Slice $\rightarrow$ Node Management \\
 & - M2 : Firewall rules running in the NaDa nodes \\
 & F5.3 - Log\_Response: Node Management $\rightarrow$ APP Slice \\
 & - M2 : Firewall rules running in the NaDa nodes \\
 & F7 - NaDa Content: NaDa Store $\rightarrow$ NaDa Management \\
 & - M12 : Encryption of channel \\
 & \hspace*{0.25cm} - M10 : Mutual authentication \\
 & \hspace*{0.25cm} \hspace*{0.25cm} - M8 : ISP Certificate \\
 & \hspace*{0.25cm} \hspace*{0.25cm} - M9 : Remote node attestation \\
 & \hspace*{0.25cm} \hspace*{0.25cm} \hspace*{0.25cm} - M6.1 : Trusted Boot, initialize measuring data for loaded code. \\
 & F8 - App\_Content: APP Slice $\leftrightarrow$ Node Management \\
 & - M2 : Firewall rules running in the NaDa nodes \\
 & F9 - App\_Content: Node Management\_2 $\leftrightarrow$ APP Slice\_2 \\
\hline
\begin{minipage}{3cm} EoP \end{minipage}
 & {\bf Processes:} \\
 & P3: APP Slice \\
 & - M1 : Mandatory access control in hypervisor (sHype) \\
 & - M2 : Firewall rules running in the NaDa nodes \\
 & - M3 : Trusted ticket for access to overlay net \\
 & \hspace*{0.25cm} - M10 : Mutual authentication \\
 & \hspace*{0.25cm} \hspace*{0.25cm} - M8 : ISP Certificate \\
 & \hspace*{0.25cm} \hspace*{0.25cm} - M9 : Remote node attestation \\
 & \hspace*{0.25cm} \hspace*{0.25cm} \hspace*{0.25cm} - M6.1 : Trusted Boot, initialize measuring data for loaded code. \\
\hline
\end{longtable}

\subsection{Abstract functional system model}\label{sec:functionalSystemModel}
As a basis for the security requirements analysis, a functional model is derived from the use
cases. The nature of the use case descriptions is such that it is not possible to identify the
complete system under investigation. Therefore an abstract functional component model is
developed, which represents the behaviour of a single node within the
system. The model 
provides an overview of every action happening at the functional borders of the node,
as well as the interactions with other nodes or with other entities of the system.

Based on the functional component model, one may now start to reason about the overall
system. The synthesis of the inner and the outer system behaviour builds the global system
behaviour. The distinction between internal and external flow description regarding the functional
component model is expressed in terms of internal and external functional
flow. This approach is based on the first results provided by the EU FP7 Evita
project and will continue the respective work from the Evita project within
NaDa. The theoretical foundation as presented in the Evita deliverable ``D3.1.2: Security and trust model''
will be applied to verify the architecture given in the previous section.

This should help to illustrate how the component model should be interpreted. Of course,
an exhaustive list of all possible instances of the system component models would be too big
to be written down. Therefore, the identification of border actions of the overall system that
are relevant to the security requirements is already performed within the component model.
However, the functional dependencies among several component instances must still be taken
into account during this process.

The work will be complemented by applying the Simple Homomorphism Verification
Tool (SHVT)~\cite{Ochsenschlaeger:Repp:Rieke:2000b}. As a first preliminary result we performed a 
high level analysis on two of the three use cases and present the results in the following sections

\paragraph{Use case 1: End User request}

\begin{figure}
\centering
\includegraphics[width=.9\textwidth]{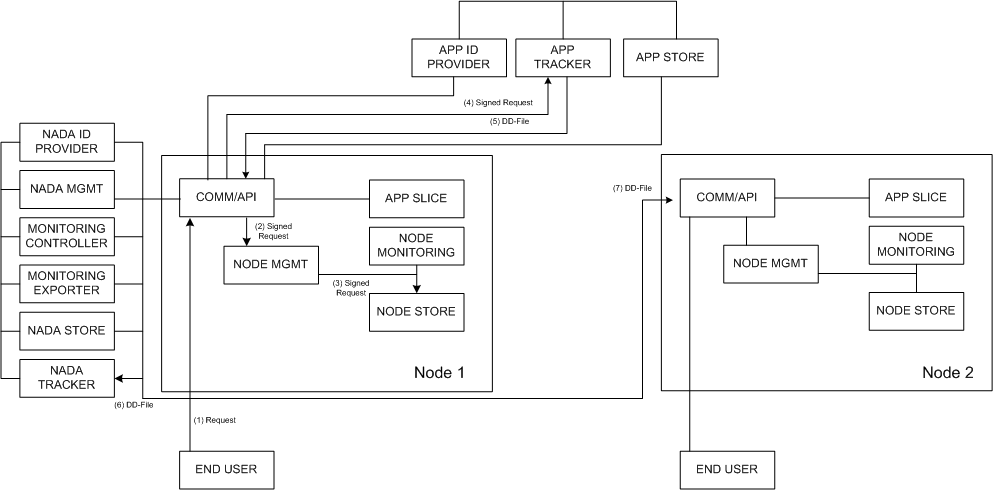}
\caption{Use case 1: End User request}
\label{fig:func1}
\end{figure}

The first diagram~\ref{fig:func1}\footnote{It is to be noted that the interaction between node 2 and the customer infrastructure is omitted for the sake of simplicity.} describes the use case scenario for end user requests. The
arrows show the communication flow and are numbered sequentially. In
contradiction to the descriptions in~\cite{kuntze:repp:2009} we consider the end of the use case with
the successful delivery of a request to the node containing the desired content.
Everything after the receipt of a delivery request is viewed by us as part of the
content delivery use case described later in this document. Note that components
that are involved in the communication are displayed gray and that for the sake
of simplicity we assume only two nodes to be involved in the process although in
distributed P2P context this is generally not the case.

The end user sends out his request, communicating with the COMM/API of the
corresponding node which is show in the diagram above with the arrow labeled with
(1). The COMM/API signs the request and sends it out to the Node Management (2)
which then stores it in the Node Store (3). After that the COMM/API sends the
signed request to the App Tracker (4) which then provides the Download Definition
File to the COMM/API (5). The COMM/API requests the NADA Tracker with the DD File
(6). The NADA Tracker sends the DD File to the node that contains the desired
content in its Slices (7) where the DD-File is processed by the corresponding
COMM/API. At this point the End User request is delivered and any reaction to the
request is assumed to be part of content distribution use case.

As the above illustrates a system of systems we can already define boundaries which are of interest. We can see that the communication that happens within a node is communication within the system. The first boundary that is crossed is arrow (4) which marks the border between node and App Tracker. The second boundary is crossed with arrow (6) between node and NADA Tracker and the third border is crossed with arrow (7) where communication happens between NADA Tracker and another node (instance) and therefore logically between two nodes (or instances).

\paragraph{Use case 2: Content Distribution}

\begin{figure}
\centering
\includegraphics[width=.9\textwidth]{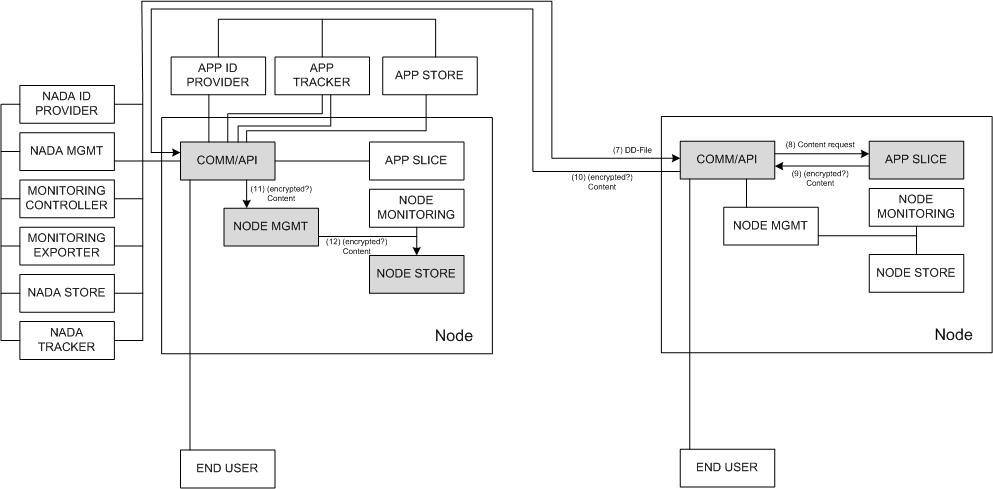}
\caption{Use case 2: Content Distribution}
\label{fig:func2}
\end{figure}

The Content Distribution, as depicted in Figure~\ref{fig:func2} begins when a DD File is received at a node in particular by the node's  COMM/API  (7).  For simplicity we assume that the desired content is available at the APP Slice of the receiving node only. Partitioned content distributed over several nodes will be described in another use case later in this content. Note that the latter case is the usual case in NADA context where the same content is available at several nodes. 
The COMM/API of the node sends the content request directly to the App Slice (8) in order to receive the desired content from it (9) the encrypted content is than sent to the COMM/API of the requesting node (10) which stores the content via Node Management (11) in the Node Store (12).

Within this use case communication takes place between the COMM/APIs of the nodes involved. These are the boundaries between the two systems (e.g. instances) which have to be considered.
%TODO Functional System Model introduction

\section{Conclusion}

This document presented the NaDa security architecture in context of the overall
system model. A threat analysis was performed identifying the security relevant
aspects and appropriate counter measures.  Next steps will include the following activities:

\begin{description}
\item[API definition] Based on the system model, abstract types, and primitives
  a API will be defined. The security implications given by the threat analysis
  are providing the frame for the development. 
\item[Implementation] Given the API a functional demonstrator will be
  implemented showing the basic security functions. For a possible code review
  the checklists presented will be used.
\item[Verification] Using the methodology from section
  \ref{sec:functionalSystemModel} leads to a model used to verify certain
  secuity relevant system functionalities.
\end{description}

\bibliographystyle{plain}
\bibliography{smv,biblio_file}

\end{document}